\documentclass[journal=ancac3,manuscript=article]{achemso}

\usepackage[version=3]{mhchem} 
\usepackage{epstopdf}
\usepackage{bm}
\usepackage{bbm}
\usepackage{mathrsfs}
\usepackage{color}
\usepackage{graphicx}
\usepackage{dcolumn}
\usepackage{amssymb}
\usepackage{amsmath}

\author{Tomas Ros\'en}
\email{trosen@kth.se}
\affiliation[FPT]
{Department of Fibre and Polymer Technology, Royal Institute of Technology, 100 44 Stockholm, Sweden}
\alsoaffiliation[WWSC]
{Wallenberg Wood Science Center, Royal Institute of Technology, 100 44 Stockholm, Sweden}
\alsoaffiliation[SBU]
{Department of Chemistry, Stony Brook University, Stony Brook, NY 11794-3400, USA}

\author{HongRui He}
\affiliation[SBU]
{Department of Chemistry, Stony Brook University, Stony Brook, NY 11794-3400, USA}

\author{Ruifu Wang}
\affiliation[SBU]
{Department of Chemistry, Stony Brook University, Stony Brook, NY 11794-3400, USA}

\author{Korneliya Gordeyeva}
\affiliation[FPT]
{Department of Fibre and Polymer Technology, Royal Institute of Technology, 100 44 Stockholm, Sweden}
\alsoaffiliation[WWSC]
{Wallenberg Wood Science Center, Royal Institute of Technology, 100 44 Stockholm, Sweden}

\author{Ahmad Reza Motezakker}
\affiliation[FPT]
{Department of Mechanics, KTH Royal Institute of Technology, SE-100 44 Stockholm, Sweden}
\alsoaffiliation[WWSC]
{Wallenberg Wood Science Center, Royal Institute of Technology, 100 44 Stockholm, Sweden}

\author{Andrei Fluerasu}
\affiliation[BNL]
{National Synchrotron Light Source II, Brookhaven National Lab, Upton, NY 11793, USA}

\author{L. Daniel S\"{o}derberg}
\affiliation[FPT]
{Department of Fibre and Polymer Technology, Royal Institute of Technology, 100 44 Stockholm, Sweden}
\alsoaffiliation[WWSC]
{Wallenberg Wood Science Center, Royal Institute of Technology, 100 44 Stockholm, Sweden}

\author{Benjamin S. Hsiao}
\affiliation[SBU]
{Department of Chemistry, Stony Brook University, Stony Brook, NY 11794-3400, USA}

\title[Exploring Nanofibrous Networks with XPCS]{Exploring Nanofibrous Networks with X-ray Photon Correlation Spectroscopy}


\begin{document}

\begin{tocentry}

\includegraphics{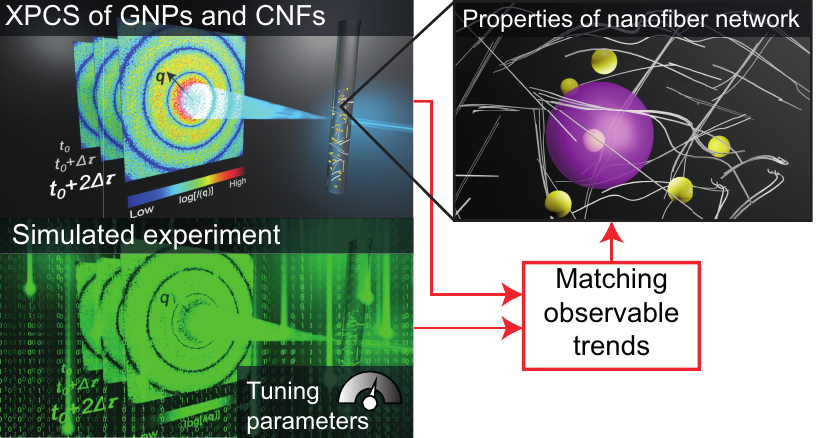}





\end{tocentry}

\begin{abstract}
Nanofibrous networks are the foundation and natural building strategy for all life forms on our planet. Apart from providing structural integrity to cells and tissues, they also provide a porous scaffold allowing transport of substances, where the resulting properties rely on the nanoscale network structure. Recently, there has been a great deal of interest in extracting and reassembling biobased nanofibers to create sustainable, advanced materials with applications ranging from high-performance textiles to artificial tissues. However, achieving structural control of the extracted nanofibers is challenging as it is strongly dependent on the extraction methods and source materials. Furthermore, the small nanofiber cross-sections and fast Brownian dynamics make them notoriously difficult to characterize in dispersions. In this work, we study the diffusive motion of spherical gold nanoparticles in semi-dilute networks of cellulose nanofibers (CNFs) using X-ray Photon Correlation Spectroscopy (XPCS). We find that the motion becomes increasingly \emph{subdiffusive} with higher CNF concentration, where the dynamics can be decomposed into several \emph{superdiffusive} relaxation modes in reciprocal space. Using simulations of confined Brownian dynamics in combination with simulated XPCS-experiments, we observe that the dynamic modes can be connected to pore sizes and inter-pore transport properties in the network. The demonstrated analytical strategy by combining experiments using tracer particles with a \emph{digital twin} may be the key to understand nanoscale properties of nanofibrous networks.
\end{abstract}

\section{Introduction}
Networks of semi-flexible nanofibers make up many key architectural elements in nature, such as actin filaments and microtubules in the cytoskeleton of eukaryote cells, collagen nanofibrils in connective tissues and bones, chitin nanofibrils in exoskeletons of arthropods, fibroin nanofibrils in silk from spiders or silkworms, and cellulose nanofibers (CNFs) in plant cell walls~\cite{gardel2004elastic,gitai2005new,jansen2018role,ling2018nanofibrils}. The primary function of the nanofibrous networks is to provide structural integrity and resist deformation from stretching, compressing, bending, and shearing forces. Additionally, these networks provide a natural porous scaffold for transport of essential substances and other nanoscale objects. Through intricate and complex hierarchical assemblies that have evolved over millions of years, the nanoscale properties of these fibrous network can be transferred to macroscale, thus enhancing the chances of survival of the species they belong to~\cite{wegst2015bioinspired}. 

Nanofibrous networks also play an important role in another modern scenario. By extracting CNFs from the plant cell wall and reassembling them in a controlled manner, there have been numerous examples of advanced biobased nanomaterials made directly from biomass. These materials possess excellent mechanical properties and functionalizability~\cite{klemm2011nanocelluloses,li2021developing}. Some examples include strong filaments for new textile materials~\cite{rosen2020elucidating,haakansson2014hydrodynamic,mittal2018multiscale,walther2011multifunctional}, membranes for water purification~\cite{voisin2017nanocellulose,sharma2020water}, aerogels for therapeutic delivery~\cite{rostami2021hierarchical}, or artificial tissues and wound dressings in the biomedical field~\cite{czaja2006microbial,hickey2019cellulose}. By letting the CNFs provide a structural scaffold, other functional nanoparticles can be introduced to make new hybrid materials that are conductive, magnetic, fire retardant, or provide structural coloring and biosensing abilities~\cite{eskilson2020self,yao2018color,hamedi2014highly,amiralian2020magnetic,wu2012ultrastrong,liu2011clay,an2018transparent}. The hydrophilic CNFs can further be dispersed in organic solvents, making them compatible as reinforcement agents in hydrophobic polymer materials~\cite{okita2011tempo,wang2019morphology}.

Despite the wide range of applications for CNF materials, the nanoscale three-dimensional structures of the CNF networks, giving rise to the material properties, are notoriously difficult to characterize~\cite{rosen2020elucidating} and therefore to control. The main techniques for characterizing CNFs include imaging techniques such as atomic force microscopy (AFM) or transmission electron microscopy (TEM), which can provide accurate descriptions of the individual CNF morphology, but fail to capture dynamics and structures in dispersions. Small-angle X-ray scattering (SAXS) can provide both statistical information about cross-section distributions~\cite{rosen2020cross} and segmental aggregation in dispersions~\cite{geng2017structure,rosen2021understanding}. Additionally, rheological studies can address the mechanical response from the network subject to external deformation~\cite{geng2018understanding}. Still, there are no characterization techniques that have been able to provide a deeper insight into the nanoscale structures of the dynamic CNF networks, and especially the nanoscale transport properties within the network. 

{\color{black} 
By \emph{in situ} measurements of the diffusion of spherical gold nanoparticles (GNPs) using X-ray photon correlation spectroscopy (XPCS), we have probed the characteristics and nanoparticle transport properties of dilute nanofibrous (CNF) networks in liquid. In the first step of the developed methodology, we characterized the morphology of CNFs using AFM and SAXS to find average nanofiber lengths, cross-sections, degree of fibrillation and segmental aggregation. Thereafter, we characterized the diffusive motion and dimensions of GNPs in pure liquid and at different CNF concentrations using XPCS and SAXS. The data was analyzed to provide the mean diffusive time scales, showing that with increasing CNF concentration and GNP size, the dynamics become: (1) slower, (2) more non-uniform, and (3) increasingly subdiffusive.

The experimental data was parametrized as dynamic modes for the case of the largest GNP, which were fitted to simulated results from a digital twin of the experiments. This made it possible to extract the physical properties, such as pore sizes and pore connectivity. Apart from providing new insights into the CNF system used in this study, the presented methodology can be readily used to quantify general nanoparticle transport properties of nanofibrous networks in liquids for various applications, \emph{e.g.}~materials science or biomedicine.
}

\begin{figure}[p]
\centering
\includegraphics[width=0.90\textwidth]{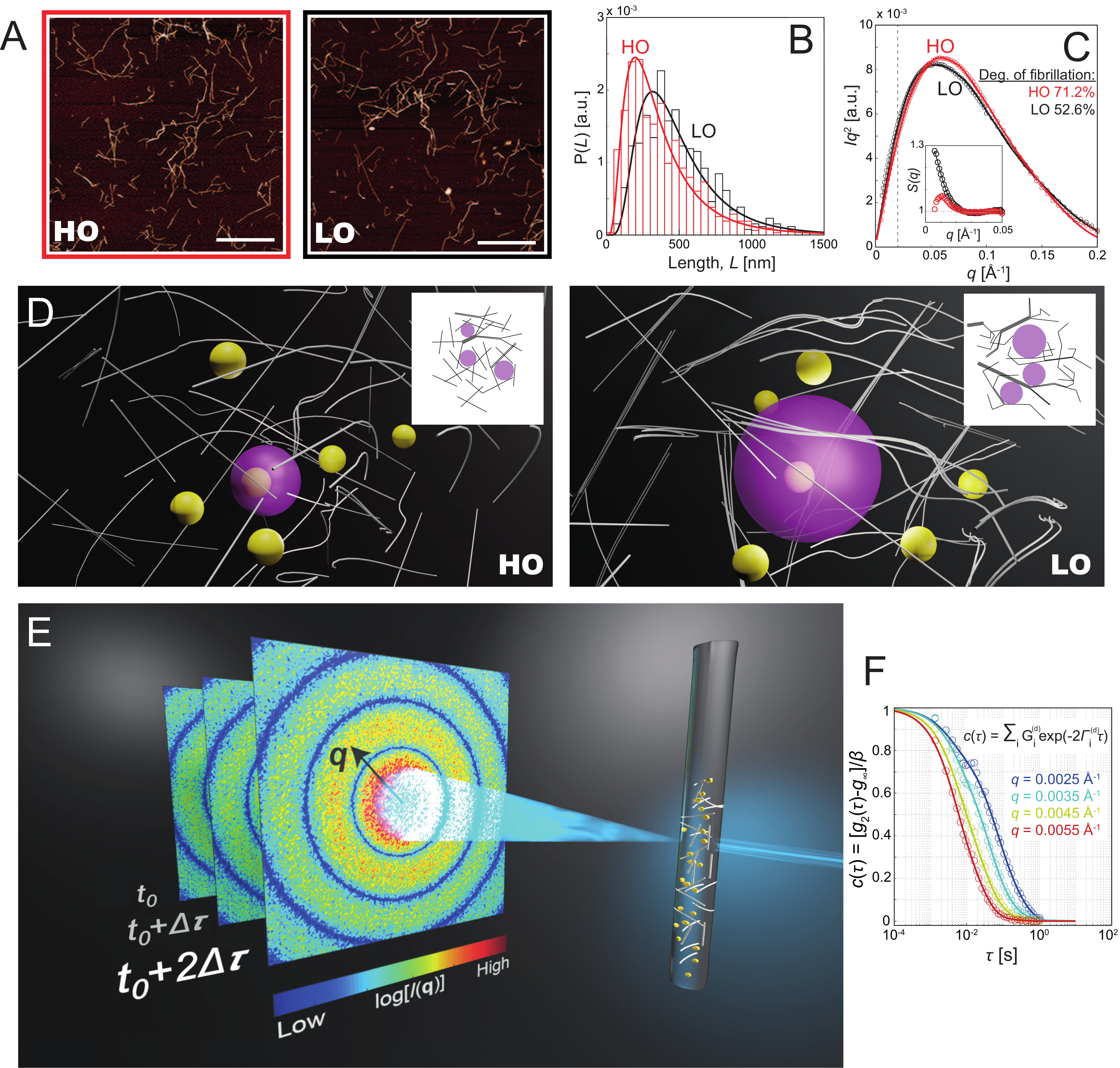}
\caption{Illustration and characterization of the samples used in this study; two different dispersions of CNFs are used with high oxidation (HO) and low oxidation (LO); (A) AFM images of typical CNFs (scale bar 1~$\mu$m); (B) length distribution of CNFs obtained from AFM images; {\color{black}(C) Lorentz-corrected SAXS data $Iq^2$ for CNFs at 0.3~wt\%, where solid curves are fits of the cross-section form factor to determine the degree of fibrillation~\cite{rosen2020cross} and the dashed line shows the lower $q$-limit for this fit; the inset figure shows the structure factor $S(q)$ of the two dispersions at low $q$}; (D) three-dimensional renderings of the two CNF networks; the higher degree of fibrillation and less local entanglements (segmental aggregates) likely result in smaller pores in the HO network (typical pore size illustrated with purple semi-transparent sphere); (E) illustration of the XPCS experiment; an X-ray beam is focused on the sample containing a mixture of CNFs and gold nanoparticles (GNPs), where the scattering intensity $I(q,t)$ is recorded on a detector and will be dominated by GNPs; (F) representative correlation curves of $c(\tau)=g_1^2(\tau)$ at various $q$ for a system with 200~nm GNPs in a HO CNF network at $0.24$~wt\%; {\color{black} the solid curves show the result from the CONTIN-algorithm}.}
\label{fig:Fig1}
\end{figure}

\section{Results}
\subsection{Morphological characterization of CNF samples}
{\color{black} Dispersions of CNFs are extracted from paper pulp typically through mechanical high-pressure homogenization.~\cite{klemm2011nanocelluloses}. Prior to homogenization, the material is usually oxidized (\emph{e.g.}~through carboxymethylation or TEMPO-mediated oxidation) to introduce charged groups, thus lowering the energy requirement for the CNF extraction due to electrostatic repulsion and allowing the CNFs to be stably dispersed in water.} The individual morphologies of CNFs are strongly dependent on extraction methods, and specifically to the degree of oxidation applied to the material prior to homogenization~\cite{rosen2020cross,geng2018understanding}. In this study, two aqueous TEMPO-oxidized CNF dispersions were prepared from wood pulp with two degrees of oxidation, which introduce different densities of charged COO$^-$ groups on the CNF surfaces and the two dispersions are referred to as HO (high oxidation) and LO (low oxidation), respectively. Through a recently demonstrated solvent exchange process~\cite{wang2019morphology}, the CNFs are stably dispersed in propylene glycol (PG). Not only is this material interesting in terms of making novel biocomposites, but it has previously been shown that the solvent exchange does not affect the CNF morphology, but rather mainly causing a reduction of Brownian motion owing to the high viscosity of PG, which will be beneficial in order to capture the dynamics with XPCS. 

Representative AFM images of the HO and LO CNFs are shown in Fig.~\ref{fig:Fig1}A. Length distributions are determined in Fig.~\ref{fig:Fig1}B, where it is found as expected that the higher oxidation treatment gives rise to shorter CNFs on average ($\bar{L} =381$~nm for HO CNFs and $\bar{L} =497$~nm for LO CNFs). {\color{black}Characterization of the dispersions at 0.3~wt\% using SAXS (Fig.~\ref{fig:Fig1}C) and fitting the Lorentz-corrected intensity curves $Iq^2$ with an assumption of quantized polydispersity~\cite{rosen2020cross} revealed that the HO CNFs also had a higher degree of fibrillation (71.2~\% compared to 52.6~\%). The inset figure in Fig.~\ref{fig:Fig1}C shows the structure factor $S(q)$ in the two dispersions, where a value of $S(q)=1$ would indicate no higher order structures, which is typical in dilute samples. The exponential shape of the LO CNF indicates strong segmental aggregation, \emph{i.e.}~locations where fibrils are locally entangled and thus in close proximity~\cite{geng2017structure,rosen2021understanding}. The structure factor is fitted to determine the average diameter of the segmental aggregates~\cite{rosen2021understanding} (solid curve in inset figure), which is found to be approximately 42~nm. The structure factor of the HO CNF instead shows a peak at $q\approx 0.01$~Å$^{-1}$ indicating that the stronger electrostatic repulsion causes nearby CNF segments to be separated at an average spacing of $d=2\pi/q\approx 60$~nm, similar to values found in dispersions of cellulose nanocrystals~\cite{rosen2021shear}.
}

The lower degree of fibrillation, longer fibrils and segmental aggregation suggest that the mass distribution inside the LO dispersion is more heterogeneous than the HO CNF, and that consequently pore sizes in the network will be larger for a given concentration (schematically illustrated in Fig.~\ref{fig:Fig1}D). Interestingly, despite these differences between the two dispersions, the rheology stays practically the same, which is shown in Fig.~S1 in the supplementary information (SI). In this work, three different concentrations of CNFs are studied, which correspond to a transition from a dilute to a semi-dilute regime: 0.08,~0.16,~and~0.24~wt\%. More details of these initial characterizations of the CNF dispersions are provided in the Materials and Methods section.

The structure of the CNF network is probed indirectly through the motion of GNPs that are added to the dispersion. The GNPs themselves are sterically stabilized with a hydrophilic polymer coating to keep them stable in electrolytes regardless of pH (more details under Materials and Methods). This also ensures negligible electrostatic interactions with the CNFs, and that they can be passively transported through the network. {\color{black} The volume fraction of the GNPs is $\Phi\approx 0.0001$ and no aggregation is observed through the structure factor of the system, thus neglecting any influence of interactions between GNPs. Three different sizes of GNPs are used here: 50,~100,~and~200~nm. Using the theory by Ogston~\cite{ogston1958spaces,chatterjee2012simple} for pore sizes in networks of randomly oriented rods (see Fig.~S2 in SI), we find that the expected pore sizes at the present CNF concentrations are of the same order as the sizes of the GNPs. Thus, we expect to be able to study the effect on the GNP dynamics from an unconfined to a confined surrounding. Three-dimensional renderings of the approximate experimental conditions in the CNF/GNP dispersions are provided in a supplementary movie.}

\subsection{XPCS: experiments, analysis and validation}
The CNF/GNP samples are mixed and injected in quartz capillaries, which are studied with XPCS at the CHX beamline (11-ID) at the National Synchrotron Light Source II (NSLS-II), Brookhaven National Laboratory, USA. The principle of XPCS is shown in Fig.~\ref{fig:Fig1}E and described in detail in the Materials and Methods section. In brief, a coherent X-ray beam (with wavelength $\lambda$) is focused on the sample and the scattering intensity $I(q)$ is recorded on a detector at various $q=(4\pi/\lambda)\sin\theta$ (with scattering angle $2\theta$). Given the much higher electron density of gold, the scattering intensity is completely dominated by the scattering from the GNPs. The interference maxima on the detector, the so called \emph{speckles}, contain information about spatial correlations between particles in the sample at a length scale $2\pi/q$. By recording a sequence of speckle patterns with delay time $\tau$, the average second-order autocorrelation of the speckle intensities over time $g_2(\tau)$ at a certain $q$ can be extracted. Through the Siegert relation, the square first-order autocorrelation is obtained by $c(\tau)=g_1^2(\tau)=(g_2-g_\infty)/\beta$, where $g_\infty$ is the baseline and $\beta$ is the speckle contrast~\cite{madsen2010beyond}. As thermal fluctuations of particle positions give rise to decaying autocorrelations of speckle intensities, the decay of $c(\tau)$ thus captures the Brownian motion in the sample. 

{\color{black}
The square first-order autocorrelation $c(\tau)$ is assumed to consist of various exponential dynamic modes $c(\tau)=\sum_i G^{(d)}_i\exp(-2\Gamma^{(d)}_i\tau)$, where the discrete distribution $G^{(d)}_i$ of log-spaced relaxation rates $\Gamma^{(d)}_i$ is found through a regularized inverse Laplace transform using the CONTIN-algorithm~\cite{provencher1982contin,andrews2018inverse} (see Fig.~\ref{fig:Fig1}F). The mean decay rate $\bar{\Gamma}$ is found through $\ln\bar{\Gamma} = \sum_{i} G^{(d)}_i\ln\Gamma^{(d)}_i$. Furthermore, a polydispersity index ($PDI$) is extracted from the relaxation rate distribution to describe the uniformity of dynamics using the method of cumulants~\cite{koppel1972analysis} (see details in the Materials and Methods section). It was also noted that the data could be well fitted with a stretched exponential function, but there are no features in the data to suggest that the decay is truly non-exponential. More discussion and details of the various fitting techniques is found in Figs.~S3 and S4 in SI.}

The potential for radiation damage was carefully assessed and in the present systems it was found that radiation was starting to influence the dynamics at a dose equivalent to 5~s of exposure from the non-attenuated beam (see Fig.~S5 in SI for details). In all the experiments in this work, we study 2 different attenuation levels for each sample: 0~\% (full beam) and 81~\% beam attenuation. As expected, no significant attenuation-dependent trends were found in the data since the total full-beam exposure never exceeded 1.4~s. Every measurement is repeated twice at a different spatial location in the sample, and we ensure that the sample is in thermal equilibrium with spatially homogeneous dynamics (see Materials and Methods section for details). {\color{black}Furthermore, any direction-dependent dynamics arising from potential gravity-induced sedimentation were examined and determined to be negligible (see Fig~S6 in SI).}

\begin{figure}[tpb!]
\centering
\includegraphics[width=0.45\textwidth]{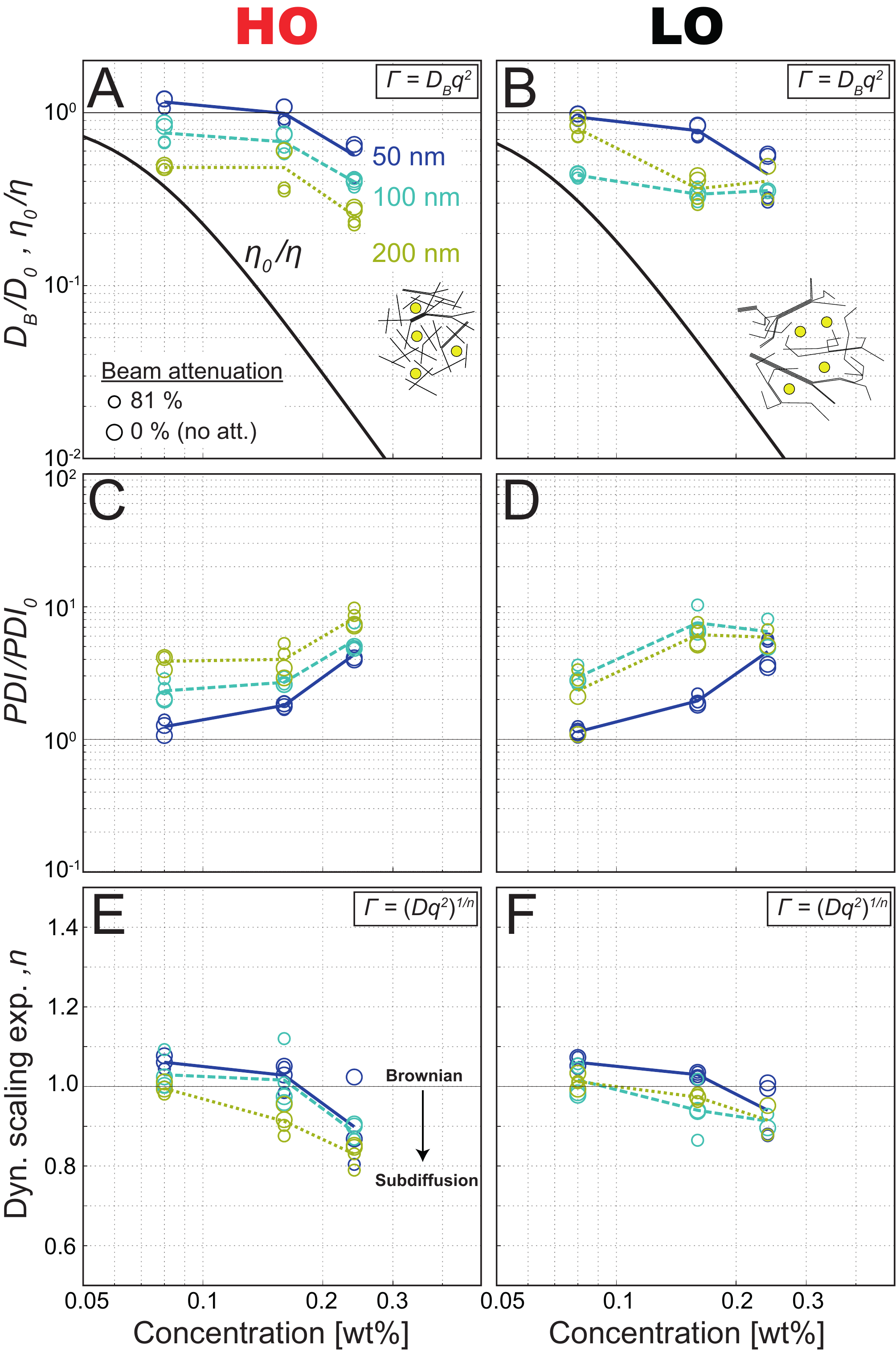}
\caption{{\color{black} Mean results from the XPCS analysis for the three sizes of GNPs; (A) and (B) show the Brownian diffusion coefficient $D_B$ of GNPs in HO and LO CNF, respectively, at various nanofiber concentrations compared to free diffusion in the solvent (propylene glycol) $D_0$ assuming $\Gamma=D_Bq^2$; the solid black curve shows the zero-shear viscosity of the CNF suspension $\eta$ compared to the viscosity $\eta_0$ of the solvent; (C) and (D) shows the polydispersity index $PDI$ as function of concentration compared to $PDI_0$ of free diffusion; (E) and (F) shows the dynamic scaling exponent $n$ assuming $\Gamma=(Dq^2)^{1/n}$, where a value of $n<1$ indicates subdiffusive dynamics. The symbol sizes correspond to the two attenuation levels used in the study.}}
\label{fig:Fig2}
\end{figure}

{\color{black} When GNPs of uniform sizes are moving freely in the solvent without CNFs, $c(\tau)$ is expected to decay exponentially (low $PDI$) with $\bar{\Gamma}=D_0 q^2$ ($D_0$ is the Brownian diffusion constant in free solvent), which was analyzed in initial experiments for the three GNP sizes. The results in Figs.~S7~and~S8 in SI illustrate that there is some inherent non-uniformity of dynamics in the GNP systems owing to a spread of actual geometric sizes (found with SAXS). Also interesting to note is that the polymer coating (with thickness $\lesssim 5$~nm) gives rise to discrepancies between geometric and dynamic radii. This effect is naturally larger for the smallest (50~nm) GNPs, where the mean geometric diameter is $39.9$~nm and mean hydrodynamic diameter is $57.5$~nm. This discrepancy does not play any major role to the study, where we investigate the \emph{relative} effect of the dynamics compared to free diffusion in the solvent,~\emph{i.e.}~the geometric diameter will stay the same but the hydrodynamic diameter will naturally change.
}

\subsection{Mean dynamics of GNPs in CNF dispersions}

Adding CNFs to the system and assuming a dispersion of straight rigid nanorods, previous theoretical works~\cite{tracy1992synthesis,kluijtmans2000self,pryamitsyn2008dynamics} suggest that the GNP motion is not hindered if the characteristic pore size (\emph{mesh size}) of the network is much larger than the GNPs, \emph{i.e.}~they will have mean diffusion constants being the same as in free diffusion $D_B=D_0$. At higher concentrations, when the mesh size is much smaller than the particle, the motion is determined by the momentum transfer within the network, which on the macroscopic scale is interpreted as an effective zero-shear viscosity $\eta$ higher than the solvent viscosity $\eta_0$. Consequently, the scaling of $D_B/D_0$ should approach the scaling of $\eta_0/\eta$. In Fig.~\ref{fig:Fig2}A, we find that the expected behavior seems to be captured fairly well in the HO CNF dispersion. Most striking is the diffusion of 50~nm GNPs with dynamics similar to free diffusion ($D_B=D_0$) below a concentration of 0.2~wt\% and then possibly approaching the scaling of $\eta$. Similar trends are seen also for the 100~and~200~nm GNPs, where of course the threshold concentration is lower with larger GNPs. Although the 50~nm GNPs show similar behavior in the LO CNF (see Fig.~\ref{fig:Fig2}B), the larger GNPs are not as affected by CNF concentration between 0.16~and~0.24~wt\%. 

{\color{black}
The mean diffusion constant however does not fully reflect the dynamic behavior in the systems. The increasing CNF concentration also results in less uniform GNP dynamics, reflected by higher values of $PDI$ compared to the $PDI_0$ in free diffusion (see Fig.~\ref{fig:Fig2}C-D). The semi-dilute CNF network thus seems to slow down the GNP motion unevenly, with both faster and slower moving particles in the system.
}

An even stronger indication that the dynamics is more complicated is found when studying the $q$-dependence of the relaxation rate $\bar{\Gamma}$ in Fig.~\ref{fig:Fig2}E-F. In a system with normal Brownian motion, the mean square displacement ($MSD$) of particles is known to be linear in time, with the diffusion constant being the proportionality constant ($MSD\varpropto D_B\tau$) with unit m$^2$/s. However, in a confined system, the dynamics can become \emph{subdiffusive} with $MSD\varpropto D\tau^{n}$, where $n<1$ and the proportionality constant gets the unit m$^2$/s$^n$~\cite{metzler2014anomalous}. To ensure this dimension of $D$, the relaxation rate is assumed the form $\Gamma=(Dq^2)^{1/n}$ and it is found in Fig.~\ref{fig:Fig2}E-F that the GNP dynamics are indeed subdiffusive in both CNF networks. Consequently, it can be argued that the comparison of $D_B$ in Fig.~\ref{fig:Fig2}A-B is not suitable, as these comparisons are based in an assumption of $n=1$. Furthermore, the values of $D$ are not comparable as their unit depends on the specific value of $n$. {\color{black} For this reason, we will now investigate the origin of this apparent subdiffusivity by studying more detailed features of the dynamic spectra and compare with a numerical simulation of the system.}

\begin{figure}[tpb!]
\centering
\includegraphics[width=0.99\textwidth]{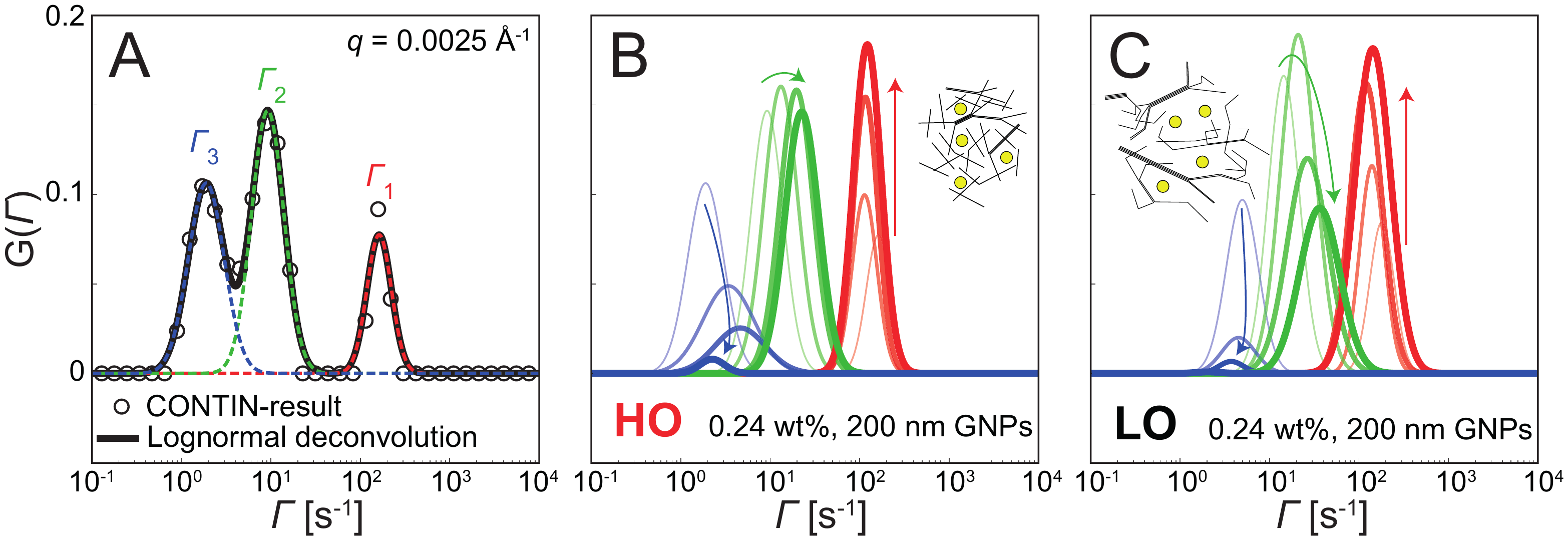}
\caption{{\color{black}Illustration of the extraction of dynamic modes using a lognormal deconvolution; (A) the resulting discrete distribution $G_i^{(d)}$ of relaxation rates $\Gamma_i^{(d)}$ at $q=0.0025$~Å$^{-1}$ as circles and the solid black line shows the continuous distribution $G(\Gamma)$ obtained by fitting three lognormal curves with mean values $\Gamma_1$, $\Gamma_2$~and~$\Gamma_3$ and magnitudes $G_1, G_2$~and~$G_3$; (B) and (C) shows the evolution of the three dynamic modes for 200~nm GNPs with increasing $q$ (illustrated with arrows) for the HO and LO CNF networks, respectively, at 0.24~wt\%.}}
\label{fig:Fig3}
\end{figure}

\subsection{Dynamic modes in the subdiffusive systems}
Although various studies have been presented recently to analyze anomalous diffusion with XPCS~\cite{pal2021anisotropic,reiser2020nanoscale,kwasniewski2014anomalous}, a different approach is employed here, similar to the method by \citet{andrews2018inverse}. {\color{black} Since Fig.~\ref{fig:Fig2} displays clear trends of slower, more polydipserse and subdiffusive dynamics with increasing CNF concentration and GNP size, we focus our attention to the system of large (200~nm) GNPs at high CNF concentration.}

{\color{black}Fig.~\ref{fig:Fig3}A shows the resulting distribution $G^{(d)}_i$ of relaxation rates $\Gamma^{(d)}_i$ from the dynamics of 200~nm GNPs in HO CNF at 0.24~wt\%, where we can distinguish three individual peaks.} Through a lognormal deconvolution, the discrete spectra is converted to a continuous function using three modes with mean relaxation rates $\Gamma_{1-3}$ and magnitudes $G_{1-3}$ as seen in Fig.~\ref{fig:Fig3}B. The evolution of these dynamic modes can then be studied individually with respect to $q$ as done in Fig.~\ref{fig:Fig3}B-C for HO and LO CNF at 0.24~wt\%, respectively. Interestingly, the individual relaxation rates $\Gamma_{1-3}$ do not monotonically increase with $q$. Instead, for both CNF samples, there is a clear switching of which modes are relevant at a given $q$-range, where the slower mode ($\Gamma_{3}$) contributes more to dynamics at low $q$, while the faster mode ($\Gamma_{1}$) contributes more at high $q$. The main difference between the two samples is that the contributions of the slow and intermediate modes ($\Gamma_{3}$ and $\Gamma_{2}$) are much lower at higher $q$.

\begin{figure}[tpb!]
\centering
\includegraphics[width=0.75\textwidth]{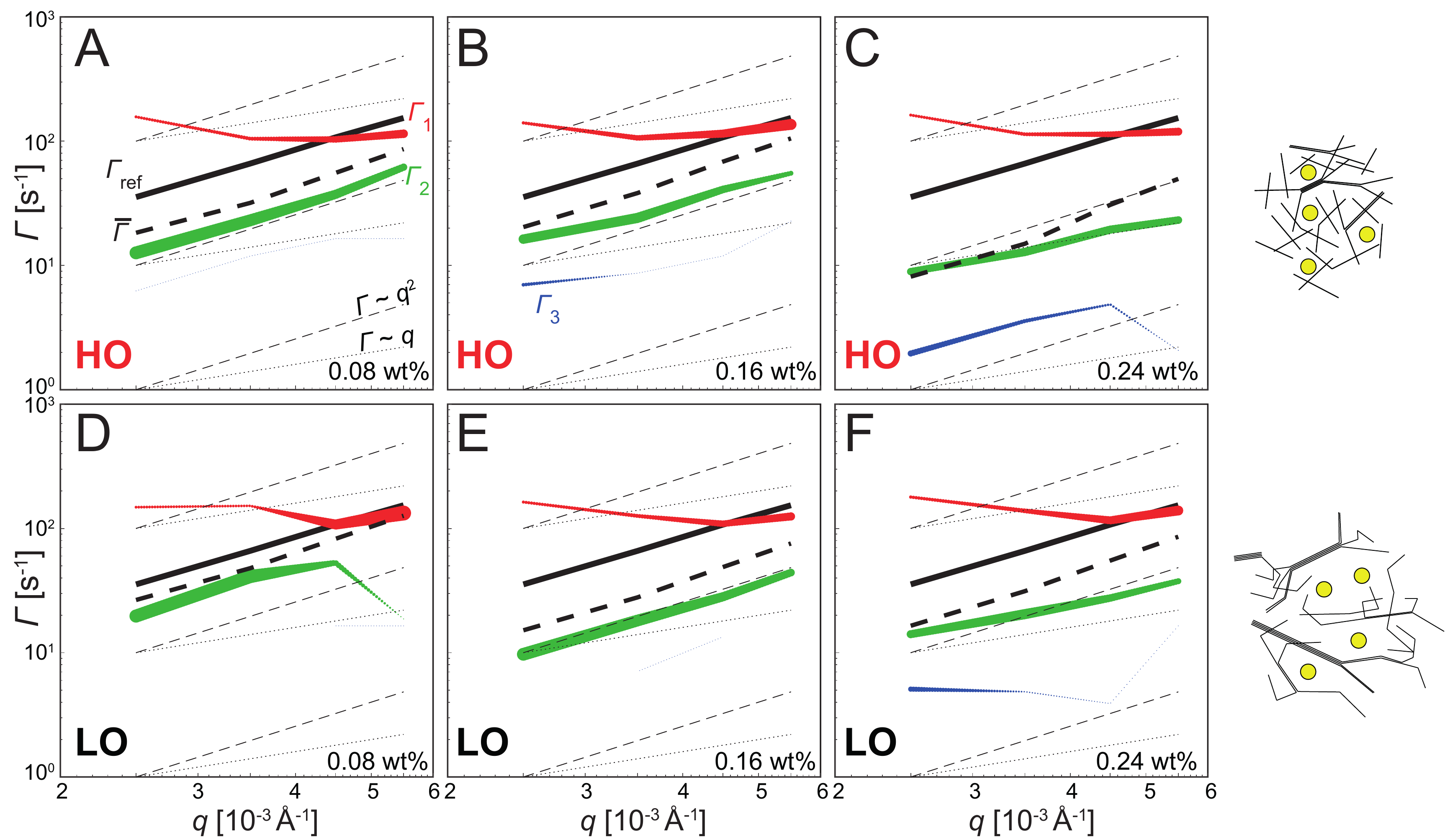}
\caption{Evolution of the relaxation rates of the dynamic modes $\Gamma_1$, $\Gamma_2$~and~$\Gamma_3$ as function of $q$ for the 200~nm GNPs; the line width is scaled with the magnitude of the mode $G_1, G_2$ or $G_3$, with linear interpolation between the $q$-values {\color{black}(note that $G_1+G_2+G_3=1$ at any given $q$)}; the thin dashed and dotted lines indicate expected scaling behavior for Brownian ($\Gamma\varpropto q^2$) and ballistic ($\Gamma\varpropto q$) dynamics, respectively; the thick black solid and dashed lines show the free diffusion of GNPs ($\Gamma_\text{ref}$) and mean relaxation rate ($\bar{\Gamma}$), respectively; (A)-(C) show the dynamic behavior in HO CNF with increasing concentration (0.08, 0.16~and~0.24~wt\%), while (D)-(E) show the behavior in LO CNF at the same concentrations.}
\label{fig:Fig4}
\end{figure}

In Fig.~\ref{fig:Fig4}, the switching of dynamic modes is illustrated in more detail. The figure shows the relaxation rates as function of $q$, where the line width of the three modes $\Gamma_{1-3}$ is scaled with their individual magnitude $G_{1-3}$. The mean relaxation rate $\bar{\Gamma}(q)$ and the reference of freely diffusing GNPs $\Gamma_\text{ref}(q)$ are plotted with thick black dashed and solid lines, respectively. The thin dashed and dotted lines correspond to the expected scaling for Brownian dynamics ($n=1$ and $\Gamma\varpropto q^2$) and ballistic dynamics ($n=2$ and $\Gamma\varpropto q$), respectively. There are some interesting similarities and differences to highlight with respect to CNF concentration in the two systems. At some low threshold concentration, there is a transition from single-mode dynamics similar to the free diffusion ($\bar{\Gamma}\approx\Gamma_\text{ref}$) to dynamics governed by two modes: (1) one faster mode $\Gamma_{1}$ that is basically constant with $q$ but only dominating at high $q$ and (2) one slower mode $\Gamma_{2}$ that still remains Brownian ($n=1$), but with slower rates than the reference and dominating at lower $q$. The main difference between the two CNFs is that the splitting of modes occurs at higher concentration in the LO CNF within the given $q$-range. In Fig.~\ref{fig:Fig4}D, for LO CNF at a concentration of 0.08~wt\%, the splitting of modes is clearly visible at around $q=0.0045$~Å$^{-1}$, which corresponds to length scales of around 140~nm. In Fig.~\ref{fig:Fig4}A on the other hand, for HO CNF at the same concentration, this mode-splitting seems to occur at higher $q$ values, but it is likely that there exist a lower concentration where the mode-splitting is at a similar $q$-value as for the LO CNF.

At higher concentrations, the faster constant mode $\Gamma_{1}$ remains at the same level as for the lower concentrations, while the slower Brownian mode $\Gamma_{2}$ decreases. However, at some concentration between 0.16~and~0.24~wt\%, it stagnates and turns towards a more ballistic behavior with $n\approx 2$ ($\Gamma \varpropto q$). At this stage, a new even slower Brownian mode appears at low $q$ ($\Gamma_{3}$). Interestingly, even though none of the contributing modes is associated with subdiffusive dynamics ($n<1$), the mean rate $\bar{\Gamma}$ is slightly subdiffusive, which is also the reason for the results presented earlier in Fig.~\ref{fig:Fig2}. \emph{The overall subdiffusive behavior is thus the result of superdiffusive modes contributing differently at different $q$.}

Additionally, since the trends with increasing concentration are similar for the two CNF networks, it is clear that the LO dispersion is going through the same stages as the HO dispersion, but generally at higher concentrations. This is of course in line with our assumption that the mass distribution of CNFs in the LO dispersion is more heterogenous and thus has a more open pore space. 

{\color{black} Although the methodology provides detailed trends of the dynamic modes in reciprocal space, it is more complicated to relate these trends to actual physical properties of the CNF network. Instead of relying on theoretical models based on idealized systems to describe the trends, we use a concept of a \emph{digital twin},~\emph {i.e.}~a numerical simulation of the system, where the same observable quantities are extracted and matched to the experiment. The same approach was recently used to study dynamics of flowing CNFs \emph{in situ} using SAXS~\cite{gowda2022nanofibril} and to study cross-sections of CNFs~\cite{rosen2020cross} with SAXS/WAXS.}

\begin{figure}[tpb!]
\centering
\includegraphics[width=0.75\textwidth]{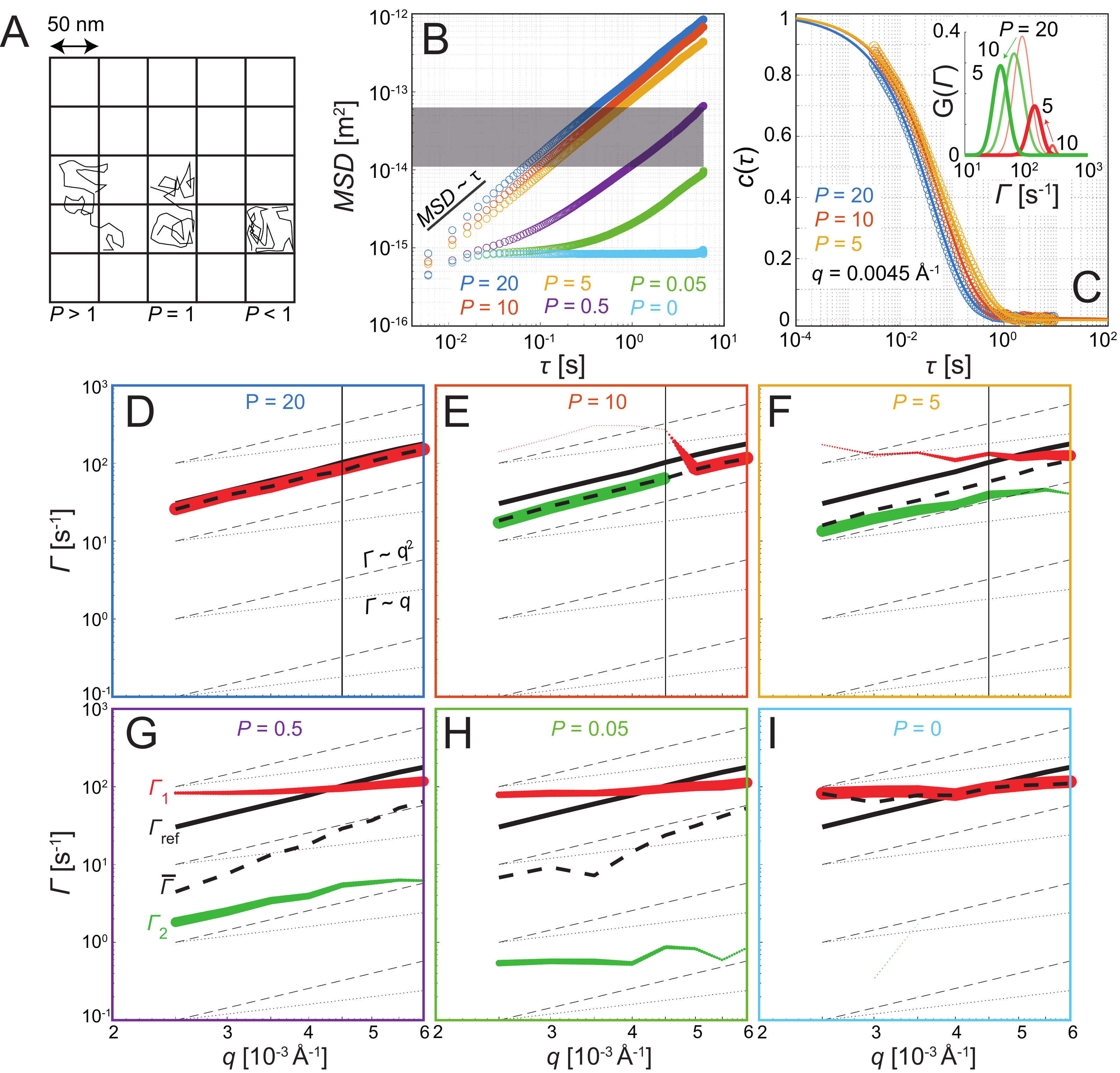}
\caption{Results from the digital twin of confined Brownian motion of 200~nm spherical nanoparticles (NP) in propylene glycol (at room temperature); (A) the simulation domain is divided into cells of size 50~nm, within which the midpoint of the NP can move freely (effective pore size is thus 250~nm); the probability for an NP to cross to another cell is determined by the permeability $P$, where $P\rightarrow\infty$ leads to free diffusion while $P=0$ leads to the NP being trapped inside the cell; (B) mean square displacement ($MSD$) as function of delay time $\tau$ and permeability $P$, where Brownian dynamics would lead to $MSD\varpropto\tau$; the shaded region shows the approximate length scales of the XPCS-experiment, \emph{i.e.} between $(2\pi/q_\text{max})^2$ and $(2\pi/q_\text{min})^2$; (C) shows the square first order autocorrelation curves $c(\tau)$ at $q=0.0045$~Å$^{-1}$ at three levels of $P$ with solid curves being the fitting result; the inset figure shows the corresponding dynamic modes after the lognormal deconvolution; (D)-(I) show the evolution of the relaxation rates $\Gamma_1(q)$ and $\Gamma_2(q)$ with increasing permeability $P$ illustrated in the same way as in Fig.~\ref{fig:Fig4}; the black line in (D)-(F) shows the $q$-value used for the data in (C).}
\label{fig:Fig5}
\end{figure}

\subsection{Development of a digital twin}
Subdiffusive motion of NPs in various semi-flexible networks has been demonstrated previously through numerical simulations, indicating that it is a natural consequence of the confined motion inside the network~\cite{metzler2014anomalous,ernst2017model,xu2021enhanced}. However, it is not clear from these models what such behavior would look like in an XPCS experiment, where the dynamics are analyzed in reciprocal space rather than real space. Therefore, in order to investigate our experimental system, we set up numerical simulations of confined Brownian spherical nanoparticles (NPs) of 200~nm, where their individual positions are used to simulate speckle patterns through a Fourier transform, which in turn can be analyzed in the same way as in the experiment. The model is practically a 2D version of the simulation method presented by \citet{ernst2017model}, where NPs are moving within a square grid providing equally sized cells, which are here chosen to have side 50~nm (see Fig.~\ref{fig:Fig5}A). Note that the cell size in the model is significantly smaller than the NP size, but since it is the midpoint that is restricted by the cell boundaries, the true pore size is actually 250~nm (the 200~nm NP can move undisturbed $\pm 25$~nm from the middle of the cell). When a Brownian step causes the midpoint of the NP to cross the cell boundary, there is a probability of passing determined by a permeability $P$ (unit s$^{-1/2}$, see Materials and Methods section for details). If the NP is not allowed to pass, it is simply bouncing off the boundary and remaining inside the cell. Thus, if $P\rightarrow\infty$, the NP is freely diffusing without any effect of the cell boundaries, but if $P\rightarrow 0$, the NP is contained within the cell where it started. Fig.~\ref{fig:Fig5}B shows the $MSD$ versus delay time $\tau$ of this system, where $P\geq 5$ demonstrates Brownian dynamics with $MSD\varpropto\tau$ ($n=1$). As the NP motion gets more and more constrained by the cell boundaries around $P=1$, the system still remains almost unaffected at small $\tau$, but is clearly subdiffusive at intermediate $\tau$ when the particle has a ''jumping'' motion from one cell to another~\cite{ernst2017model,xu2021enhanced}. When $P=0$, the value of $MSD$ is naturally limited by the size of the cell and remains constant for all larger $\tau$.

In reciprocal space, at $q=0.0045$~Å$^{-1}$, after performing the same dynamic mode analysis, we find very similar behavior as in the experiment (see Fig.~\ref{fig:Fig5}C). The dynamic behavior at $P=20$ clearly is described by one single dynamic mode, with a faster mode appearing as the permeability decreases. The entire evolution of the splitting of modes with decreasing permeability is illustrated in Figs.~\ref{fig:Fig5}D-I, which looks almost identical to the trends found in the experimental data. From being almost indistinguishable from free diffusion at $P=20$, the dynamic behavior is described with two separate modes at lower $P$, where the faster mode remains stationary both with $q$ and with decreasing $P$. The slower mode quickly slows down further with lower permeability, and with lower and lower contribution at higher $q$. Just like in the experiments, although both modes are superdiffusive with $n>1$, the mean relaxation rate still shows a subdiffusive behavior (thick dashed line). Eventually, as the contribution of the slower mode diminishes, the dynamic behavior at $P=0$, where NPs are completely contained inside the cell, is dominated by the fast stationary mode. The two modes can thus be explained by the two types of NP dynamics at different length scales using this simplified model: (fast) inter-cell dynamics and (slow) macroscopic "jumping" dynamics between cells.

\begin{figure}[tpb!]
\centering
\includegraphics[width=0.75\textwidth]{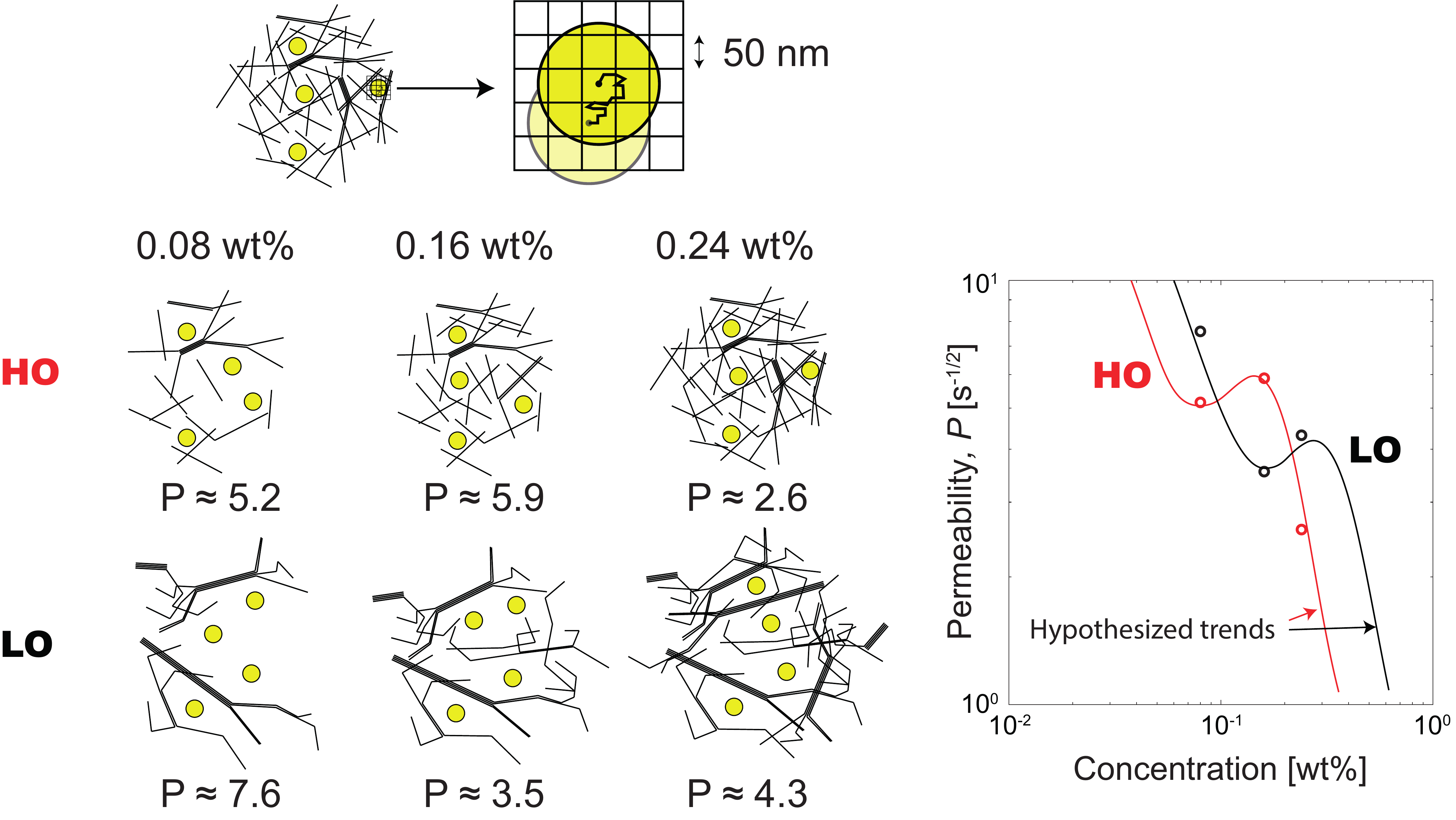}
\caption{{\color{black} Schematic illustration of the dynamics of 200~nm GNPs in the two CNF dispersions during network formation with increasing concentration, quantified by permeability $P$ by matching dynamic features to the digital twin; at dilute conditions, the GNPs are moving freely within the solvent and are unaffected by the fast Brownian CNFs; as concentration reaches a level where the average pore sizes are smaller than the GNPs, a local cage structure of nanofibers will restrict the GNP motion, but it has some probability of moving due to structural changes of the dynamic network; the local structure and dynamics would be the source of the apparent permeability $P$ and unrestricted wiggle-room of $\pm 25$~nm; the combination of less dynamic reconfiguration, and smaller pore sizes at higher concentrations gives rise to a plateau of the permeability.}}
\label{fig:Fig6}
\end{figure}

\subsection{Discussion}
How can the above observation be connected to the network properties of the two CNF dispersions in the experiment and what do the simulated cell size and permeability correspond to for the real network structure? The GNPs undergo Brownian motion that is determined mainly by the interaction with solvent molecules. Individual CNFs are also highly Brownian with much faster dynamics than the GNPs, which can be interpreted on a nanometer scale as a solvent with seemingly higher apparent viscosity, thus slowing down the motion of the GNPs. At a higher CNF concentration, when the average pore becomes smaller than the GNP size, a loose cage of CNFs is also formed around the individual GNPs. The cage is naturally slightly larger than the GNP and provides some restriction to its motion. {\color{black}However, the GNP can escape its local cage as the surrounding dynamic CNF network will constantly undergo structural changes causing escape routes for the GNP.} This probability of the GNP escaping its local cage could thus be perceived as the source of the permeability in the model. With an increasing CNF concentration, the cage will become stronger and GNPs will remain essentially stationary, which corresponds to the case of zero permeability. 

{\color{black} The main permeability-dependent feature of the dynamic modes in Fig.~\ref{fig:Fig5}, is the decrease of the second mode $\Gamma_2$ with decreasing $P$, which also seems to correspond to the effect observed in the experimental system. By matching the mean rate of $\Gamma_2$ between the experiment and the digital twin, we can obtain quantitative values of $P$ in the experimental system as illustrated in Fig.~\ref{fig:Fig6} (see details in Fig.~S10 in SI). At low concentrations the permeability is higher in the LO CNF compared to HO CNF at a given concentration, which makes sense owing to the heterogeneous mass distribution and therefore less direct interactions for the GNPs with the CNFs. Interestingly, both for the HO CNF and LO CNF, we observe a slight increase of permeability values in a certain concentration range. We hypothesize that this is linked to the actual network formation, where the Brownian dynamics of CNFs slow down rapidly, thus decreasing the CNF-GNP interactions despite the smaller pore sizes. When the network is formed, the permeability $P$ again decreases rapidly with concentration. Somewhat counter-intuitive, this plateau of $P$ occurs at higher concentrations for LO CNF suggesting that the heterogeneity of the dispersion causes network formation at a higher volume fraction despite the larger CNF aspect ratios. At high concentrations, the LO CNF will again have higher permeability than the HO CNF owing to the larger physical pore sizes in the network.}

It is natural to assume that the model cell size (or rather wiggle-room) should also be linked to CNF concentration. However, doing the same analysis for both larger (75~nm) and smaller (25~nm) model cell sizes reveals a different scenario for the splitting of modes (see Fig.~S9 in SI for details), which indicates that both CNFs in our study indeed create a local surrounding of the 200~nm GNP with an unrestricted wiggle-room of $\approx 50$~nm. We hypothesize that this apparent wiggle-room might rather be influenced by the flexibility of individual CNFs. The flexibility might also be influencing the magnitude of the stationary mode ($\Gamma_1$). In a recent similar XPCS study by \citet{reiser2020nanoscale}, the dynamics of spherical NPs in networks of wormlike micelles also revealed a stationary mode, which was argued to be related to the Kuhn length of the individual micelle. This could naturally not be investigated in this work as the cell boundaries are fixed.

Although revealing much of the relevant dynamics in the experiment, there are some features that the reduced model cannot capture. For example, the emergence of a third dynamic mode ($\Gamma_3$), which suggests the real system also includes dynamics on a higher hierarchical level. Obviously, the "network" in the reduced model is static, while the network of CNFs is dynamic, and it is therefore intuitive to assume that there should be an additional dynamic mode connected to the network dynamics, although this requires further investigation.

\section{Conclusions}
This work addressed the dynamics of sterically stabilized spherical gold nanoparticles (GNPs) in semi-dilute dispersions of cellulose nanofibers (CNFs) using X-ray photon correlation spectroscopy (XPCS). CNFs were prepared at two different degrees of oxidation, which provided differences in the CNF length distribution as well as the degree of fibrillation, which in turn had a clear effect on the pore spaces in the network at a given concentration. From the smallest GNPs (50~nm), it is found that their dynamics are unaffected by CNFs up until a concentration of $\approx 0.2$~wt\% in both systems, suggesting that the undisturbed pore sizes in the semi-dilute networks were of the similar order. However, the dynamics of the larger GNPs (200~nm) could be described by several dynamic modes, which through a comparison with a digital twin were found to be connected to the GNPs' ability to escape their local surrounding of CNFs. {\color{black} This in turn could be used to quantify the effective permeability of the GNPs in the dynamic semi-dilute CNF network and to characterize threshold concentrations for network formation.}

The fact that the relatively simple model of confined Brownian motion was able to capture the main dynamic modes from the demonstrated experiment in this study opens up vast possibilities for exploring nanofibrous networks by characterizing the dynamics of tracer NPs using XPCS. The obvious next step would be to study a wider $q$-range with the digital twin and carry out a proper parameter study of both pore size and permeability to pinpoint how these parameters can affect the measurable quantities. {\color{black}This in turn can aid in designing new improved experiments. As a future outlook, the digital twin can be improved using coarse-grained molecular dynamics simulations of a realistic nanofibrous network to directly connect the dynamic modes to CNF properties and interactions, possibly also capturing features that cannot be measured experimentally. This could potentially also be used as training data for machine learning, allowing physical properties to be directly obtained from the experimental data.

Finally, we would like to emphasize that the methodology is not limited to our experimental system but can be used more generally to other nanofibrous networks, \emph{e.g.} from polymeric, ceramic or carbon fiber networks. Technically, the same methodology could also be applied to systems without a bounding solvent, \emph{e.g.} transport in membranes or aerogels, although it might me more challenging/hazardous experimentally to study the unbounded NP transport through these networks.
}

\section{Materials and Methods}
\subsection*{Samples}
The spherical gold nanoparticles (GNPs; diameters of 50~nm, 100~nm and 200~nm) dispersed in propylene glycol were purchased from Nanopartz\texttrademark~and were functionalized with a proprietary hydrophilic polymer with a MW~1500~Da (product name H11-PHIL-PRG). The dispersion was characterized by the vendor using both transmission electron microscopy (TEM) and dynamic light scattering (DLS) ensuring the quality and uniformity of the GNPs. Analysis in the present work using SAXS further confirmed high uniformity of the GNP sizes (see Figs.~S7 in SI for details).

Never dried sulfite softwood pulp (Domsjö dissolving pulp) was provided by RISE Research Institute of Sweden and was used for preparation of cellulose nanofibers (CNFs). Sodium hypochlorite ($6$-$12$~wt\%, Alfa Aesar), TEMPO (2,2,6,6-tetramethyl-1-piperidinyloxy free radical, $\geq98$~\%, Alfa Aesar), sodium hydroxide ($\geq99.2$~\%, VWR Chemicals), sodium bromide (BioUltra, $\geq99.5$~\%, Sigma Aldrich), concentrated hydrochloric acid (VWR Chemicals, $\geq35$~\%) were used as received. The TEMPO-mediated oxidation was performed on never dried sulfite softwood pulp (41.6~wt\%) following the protocol described elsewhere~\cite{saito2007cellulose}, where two degrees of oxidation were achieved by adding 3 and 1.5~mmol of oxidant per gram of pulp for high oxidation (HO) and low oxidation (LO) CNFs, respectively. To extract CNFs, treated pulp was fibrillated by a high pressure (1600~bars) Microfluidizer (M-110EH, Microfluidics) with a 400/200~$\mu$m (1 pass) and a 200/100 $\mu$m (4 passes) wide chambers connected in series. The obtained 1~wt\% CNF gel was further diluted to 0.3~wt\% and homogenized by Ultra-Turrax dispersing tool (IKA, Sweden) for 10~min at 10000~rpm. Conductometric titration was performed to estimate the total CNF charge according to protocols described elsewhere~\cite{katz1984determination} and average charge densities of carboxylic (COO$^-$) groups were estimated to be $448\pm16$ and $1023\pm26$~$\mu$mol/g for the LO and HO CNF, respectively. The solvent exchange process to propylene glycol (PG) followed the same procedure as described elsewhere~\cite{wang2019morphology}, and the resulting PG-CNF dispersions were used to create dilutions of $0.1, 0.2$~and~0.3~wt\%. The final CNF concentrations $0.08, 0.16$~and~0.24~wt\% are reached after mixing with the GNP dispersions. 

The mixing of GNPs into the CNF dispersions resulted in a final GNP concentration of 0.2~wt\%, which is comparable to the weight fraction of CNFs. However, since the volume of a single GNP is much larger than a CNF, the number density in the network is still quite low (see examples of 3D rendered networks in the supplementary movie), and thus not affecting the overall structure of the network. The samples were mixed and dispersed through vortex-mixing (VWR Fixed -Speed Mini) and remain static for 24~hours before the experiment. Before the actual XPCS measurements, the samples were re-dispersed by vortex-mixing (ORIGINAL Vortex-Genie 2) and then put into an ultrasonic bath for 5~minutes to prevent precipitation of nanoparticles. The samples were then injected to thin-walled quartz capillary tubes for the XPCS experiments with an approximate time of 1-2~h between injection and measurement.

\subsection*{Atomic Force Microscopy (AFM)}
Topography of particles was visualized with the help of an Atomic Force Microscope (AFM, Multimode 8, Bruker, USA) in Tapping mode in air. The original dispersion of CNFs (0.3~wt\% in DI water or PG) was diluted to 0.005~wt\% and homogenized for 10~min at 10000~rpm using Ultra Turrax (IKA T25D, Sweden). Remaining aggregates were removed by centrifugation for 1~hour at 4000~rpm. Supernatant was collected and mixed on Vortex Genie~2 (Scientific Industries Inc., USA) for 5~min right prior casting. A 20~$\mu$L drop of (3-aminopropyl)thriethoxysilane (99\%, Sigma Aldrich) was placed and kept on a freshly cleaved mica for 30~s, whereafter it was blown away intensively by nitrogen gas. A 20~$\mu$L drop of CNF dispersion (0.005~wt\%) was placed rapidly on functionalized mica and kept for 30~s until being blown away by nitrogen gas in a similar manner. Obtained substrates were left to dry overnight. Diameters and lengths of CNFs were estimated from height images using software Nanoscope Analysis and ImageJ, acquiring above 300~and~500 measurements, respectively. With the help of software Origin 2021 statistical analysis was performed on collected values and by fitting a Gaussian distribution function it was estimated that average diameters (dav) for high and low charge CNF were $2.73\pm0.63$ and $2.56\pm0.51$~nm, respectively. Both CNF types demonstrated a broad distribution of lengths varying from 50~nm~to~1.6~$\mu$m. By fitting a lognormal distribution, the mean lengths $\bar{L}$ were determined.

\subsection*{Small-angle X-ray scattering (SAXS)}
The SAXS experiments were performed at the LIX beamline (16-ID) at the National Synchrotron Light Source II (NSLS-II), Brookhaven National Laboratory, USA. The CNF samples (without GNPs) are injected into liquid sample holders with mica windows, which are scanned at five different positions with 1~s exposure at each position, where a mean was used for analysis. The wavelength was $\lambda = 0.82$~Å and the sample-detector distance was 3.7~m. The beam size was approximately 50~$\times$~50~$\mu$m$^2$. The scattered X-ray intensity $I(q)$ was recorded on a Pilatus 1 M detector with pixel size 172~$\times$~172 $\mu$m$^2$ at a range of $q = 0.006 - 0.2$~Å$^{-1}$, with scattering vector magnitude $q = (4\pi/\lambda)\sin\theta$ (the angle between incident and scattered light is 2$\theta$). The background scattering intensity was subtracted using scans of DI water. The degree of fibrillation of the CNFs is determined with the fitting method by \citet{rosen2020cross} {\color{black} applied to the Lorentz-corrected SAXS data $Iq^2$ between 0.02~and~0.2~Å$^{-1}$, resulting in the form factor of the system $P(q)$. The structure factor is found through $S(q)=I(q)/P(q)$, where the radius of segmental aggregates $R$ can be found by fitting $S(q)=A\exp[-(0.396R)^2q^2]+1$ between 0.005~and~0.05~Å$^{-1}$ according to previous work\cite{rosen2021understanding}.}

\subsection*{Rheology}
Steady shear viscosity measurements were performed on Rheometer DHR-2 (TA Instruments, USA) using plate-on-plate geometry (diameter 25~mm) with the gap set to 1~mm. Prior to the measurements, the CNF dispersions were degassed by vacuum to remove air bubbles and left to rest at normal conditions for at least overnight. The experiment consisted of two steps: sample conditioning at 25~$^{\circ}$C for 5~min and flow sweep test with shear rate varying between 0.1~and~100~s$^{-1}$. Data was acquired in a logarithmic manner collecting 10~points per decade with equilibration time set to 5~s and averaging time reaching 30 s. 

The approximate zero-shear viscosity $\eta$ of the CNF dispersions was obtained by fitting the experimental curves. More details about the rheological characterization is provided in Fig.~S1 in SI.

\subsection*{X-ray photon correlation spectroscopy (XPCS)}
The XPCS experiments were performed at the CHX beamline (11-ID) at NSLS-II, BNL. An X-ray beam with size 40~$\times$~40~$\mu$m$^2$ and wavelength $\lambda = 1.28$~Å was focused on the sample and the scattered X-rays were collected on a detector (Dectris 3-Eiger X, pixel size 75~$\times$~75 $\mu$m$^2$) at a sample-detector distance of 16~m covering $q = 0.002 - 0.04$~Å$^{-1}$. Each sample was scanned with 1000~images at three different exposure times (1.3,~10~and~100~ms; also setting the frame rate) and four attenuation levels (0, 81, 96.4~and~99.3~\%). Each scan was repeated twice, yielding a total of 24 measurements per sample, which all were taken at different positions in the capillary. It was however found that only the data with the lowest exposure time of 1.3~ms (frame rate 750~Hz) could be used for analysis and the scattering intensity using the highest attenuation levels (96.4 and 99.3~\%) was too low to be used for standard XPCS analysis. An assessment of the influence of the beam on the sample dynamics revealed that the samples were affected with a total dose equivalent to 5~s of exposure to the non-attenuated beam (see Fig.~S5 in SI for details), which is approximately four times higher than the maximum dose in any experiment used for analysis. The auto-correlation $g_2(q,\tau)$ at a certain delay time $\tau$ of the intensity at a certain pixel $I(q,t)$ is determined both through the one-time correlation function (1T) {\color{black}obtained through a multi-tau algorithm~\cite{lumma2000area}}:\\
\begin{equation}
g_2^{(1T)}(q,\tau)=\frac{\langle I(q,t) I(q,t+\tau)\rangle}{\langle I(q,t)\rangle^2}
\end{equation}
~\\
or through the mean over all $\tau = |t_2-t_1|$ of the two-time correlation function (2T)~\cite{fluerasu2007slow}:\\
\begin{equation}
g_2^{(2T)}(q,\tau)=\left\langle\frac{\langle I(q,t_1)I(q,t_2)\rangle}{\langle I(q,t_1)\rangle\langle I(q,t_2)\rangle}\right\rangle_{\tau = |t_2-t_1|}.
\end{equation}
~\\
The average correlations were determined in circular regions on the detector with centers $q=0.0025, 0.0035, 0.0045$~and~$0.0055$~Å$^{-1}$, where each region had a width of $\Delta q=0.001$~Å$^{-1}$. Ideally, the two definitions above should provide the same results, but it was found that they differ for low scattering samples {\color{black}because of spurious parasitic scattering.} Here, a data set was discarded due to low scattering intensity if the average relative difference ($ARD$) between $g_2^{(1T)}$ and $g_2^{(2T)}$ was larger than 1~\%. Otherwise, a mean of the two was used as the true auto-correlation $g_2(q,\tau)$. To ensure spatial homogeneity of the samples, the repeated scans (at different positions) were used to determine the mean $g_2(q,\tau)$. However, if one repeated experiment showed significantly higher correlations than its counterpart ($ARD>0.5$~\%), it was considered an outlier and removed from the analysis. The final square first order auto-correlation used for analysis $c(q,\tau)=g_1^2(q,\tau)=(g_2(q,\tau)-g_\infty)/\beta$ is found by subtracting the baseline $g_\infty=g_2(q,\tau>5~s)$ and scaling with the speckle contrast $\beta$ ($\beta = \beta^*+1-g_\infty$, where $\beta^*$ is found through initial calibration with a test membrane). {\color{black} The time-dependence of the dynamics over the course of each experiment was also assessed with the two-time correlation function to exclude any heterogeneous dynamics or other time-dependent phenomena.}

{\color{black}
\subsection*{XPCS analysis}
Each resulting correlation curve $c(\tau)$ at a certain $q$ was assumed a form of:\\ 
\begin{equation}
c(\tau)=\sum_i G^{(d)}_i\exp(-2\Gamma^{(d)}_i\tau), 
\end{equation}
~\\
where the discrete distribution $G^{(d)}_i$ is found through a regularized inverse Laplace transform using the CONTIN-algorithm implemented in MATLAB 2020b and adapted from the code by \citet{MATLAB_RILT}. Here, we assumed 50 log-spaced modes in range $\Gamma^{(d)}_i\in[75,750000]$~s$^{-1}$ with constraints $G^{(d)}_i\geq0\quad\forall i$, $G^{(d)}_i\rightarrow 0$ at the extreme values of $\Gamma^{(d)}_i$ and that $\sum_i G^{(d)}_i =1$, \emph{i.e.} that $c(\tau\rightarrow 0)=1$. The CONTIN-algorithm requires a regularization parameter for additional smoothing of the distribution~\cite{provencher1982contin,andrews2018inverse}, which was here chosen to 0.05. However, an assessment of the influence of the regularization parameter (provided in Figs.~S3 and S4 in SI) showed that the exact value had little effect on the extracted dynamic modes for $RP\leq 0.1$.

The mean relaxation rate $\bar{\Gamma}$ of the decay is determined from the distribution through:\\
\begin{equation}
\ln\bar{\Gamma} = \sum_{i} G^{(d)}_i\ln\Gamma^{(d)}_i.
\end{equation}
~\\
The $q$-dependence of $\bar{\Gamma}$ was then used to find the Brownian diffusion coefficient $D_B$ through fitting $\bar{\Gamma}(q)=D_B q^2$ and the dynamic scaling exponent $n$ through $\bar{\Gamma}(q)=(Dq)^{1/n}$.

Through a cumulant expansion~\cite{koppel1972analysis}, the initial decay ($\tau\rightarrow 0$) of the auto-correlation can be approximated with $\ln c(\tau)\approx-2\bar{\Gamma}_0\tau+k_2^2\tau^2+O(\tau^3)$, where $\bar{\Gamma}_0$ is the mean relaxation rate at $\tau=0$ (note that $\bar{\Gamma}_0\neq\bar{\Gamma}$). The polydispersity index ($PDI$) is determined through $PDI=k_2^2/\bar{\Gamma}_0^2$. By doing a low $\tau$ expansion of $c(\tau)$ both from the CONTIN result and the cumulant expression and comparing terms, it can be found that:\\
\begin{eqnarray}
\sum_i G_i^{(d)}&=&1\\
\sum_i G_i^{(d)}\Gamma_i^{(d)}&=&\bar{\Gamma}_0\\
\sum_i G_i^{(d)}\left(\Gamma_i^{(d)}\right)^2&=&\left(\bar{\Gamma}_0^2 + \frac{k_2^2}{2}\right)
\end{eqnarray}
~\\
and the polydispersity index can thus be found from the discrete relaxation rate distribution $G^{(d)}_i$:\\
\begin{equation}
PDI = \frac{2\left(\sum_i G_i^{(d)}\left(\Gamma_i^{(d)}\right)^2 - \bar{\Gamma}_0^2 \right)}{\bar{\Gamma}_0^2 }.
\end{equation}
~\\
The polydispersity index was seen not seen to vary significantly with $q$ and a mean value was taken over all $q$ for comparisons in Fig.~\ref{fig:Fig2}. The relative quantities $D_B/D_0$ and $PDI/PDI_0$ in Fig.~\ref{fig:Fig2} were obtained by using the values $D_0$ and $PDI_0$ from experiments of free diffusion of GNPs in the solvent (see Fig.~S8 in SI for details).

\subsection*{Extraction of dynamic modes through lognormal deconvolution}
The lognormal deconvolution is used to create a continuous distribution $G^{(c)}(\Gamma)$ of the resulting discrete distribution $G^{(d)}_i$ and was done by assuming the form with three dynamic modes in the continuous distribution ($\Gamma_1$, $\Gamma_2$,~and~$\Gamma_3$):\\
\begin{equation}
G^{(c)}(\Gamma) = \sum_{k=1}^3 G^{(c)}_k \exp(-a_k[\ln(\Gamma)-\ln(\Gamma_k)]^2),
\end{equation}
~\\
where magnitude $G^{(c)}_k$, relaxation rate $\Gamma_k$ and parameter $a_k$ of each mode $k=1,2,3$ are found through fitting. The fitted magnitude $G^{(c)}_k$ is then used to determine the contribution of each mode $G^{(d)}_{k,i}$ to the original discrete distribution $G^{(d)}_i$, where $G_i^{(d)}=\sum_{k=1}^3 G^{(d)}_{k,i}$. The magnitudes of each mode $G_k$ is found by $G_k = \sum_i G^{(d)}_{k,i}$, which fulfills $\sum_{k=1}^3 G_k=1$.
}

\subsection*{Numerical simulations of confined Brownian motion}
The numerical simulation model is a 2D version of the model presented by \citet{ernst2017model} to simulate subdiffusive Brownian motion in mucus. Here a square domain of $L=5$~$\mu$m was divided into $n=L/l$ quadratic subdomains (''cells'') with cell size $l=50$~nm (analysis of cell sizes $l=25$~\&~75~nm is provided in Fig.~S9 in SI), where $N=100$ particles are initially randomly placed in the domain. The position of a particle, $\boldsymbol{r}(t)$ equals to integration of the stochastic differential equation (SDE) $\dot{\boldsymbol{r}}(t)=\sqrt{2D_0}\boldsymbol{\zeta}(t)$, where $\boldsymbol{\zeta}(t)$ is a stationary Markov process satisfying $\langle\boldsymbol{\zeta}(t)\rangle=\boldsymbol{0}$. The diffusion constant is set according to $D_0=k_BT/(6\pi\eta R)$, where radius is $R=100$~nm, temperature is $T=293$~K, and viscosity is $0.042$~Pa~s (viscosity of PG at this temperature~\cite{sagdeev2017density}). If a particle crosses a boundary of a cell, it will be reflected with probability $r=1-P\sqrt{\Delta\tau}$, where $P>0$ is the membrane permeability with unit s$^{-1/2}$ and $\Delta\tau=1$~ms is the time step. The simulation is running for a total of 6000 iterations (\emph{i.e.}~6~s). The mean-square-displacement is calculated as the ensemble average over all particles $MSD(\tau)=\langle[\boldsymbol{r}(t=\tau)-\boldsymbol{r}(t=0)]^2\rangle$. It can be noted that in the extreme case of $P=0$, the particle is contained within its pore with $MSD(\tau\rightarrow\infty)=l^2/6$~\cite{ernst2017model}.

\subsection*{Simulated XPCS experiments}
The simulated speckle patterns were created by a pre-defined detector with $200 \times 200$ pixels covering a range of $q_x,q_y\in[-0.0065,0.0065]$~Å$^{-1}$. The dynamic structure factor at a given time $S(q,t)$ was calculated using a discrete Fourier transform of the particle positions $\boldsymbol{r}(t)$:\\
\begin{equation}
S(q,t)=\left|\sum_{i=1}^N \exp[j\boldsymbol{q}\cdot\boldsymbol{r}_i(t)]\right|^2,
\end{equation}
~\\
with $j=\sqrt{-1}$. The intensity $I(q,t)=P(q)S(q,t)$ is found by multiplication of the form factor for a sphere~\cite{feigin1987structure}:\\
\begin{equation}
P(q)=\left(\frac{\sin(qR)-qR\cos(qR)}{(qR)^3}\right)^2
\end{equation}
~\\
using radius $R=100$~nm. The $q$-regions for evaluating intensity autocorrelations in the simulated detector images were chosen with centers $q=0.0025,0.003 ... 0.006$~Å$^{-1}$ and width of $\Delta q=0.005$~Å$^{-1}$. The resulting autocorrelations are then analyzed through the same steps as the experimental data by deriving the square first order correlation $c(\tau)=g_1^2(q,\tau)=(g_2(q,\tau)-g_\infty)/\beta$, where $g_2(q,\tau)=g_2^{(2T)}(q,\tau)$ and $\beta$ is found through extrapolation to $\tau=0$.

\begin{acknowledgement}
The authors acknowledge the financial support from the National Science Foundation (DMR-1808690), Wallenberg Wood Science Center (WWSC), the Alf de Ruvo Memorial Foundation and the Hans Werthen Foundation. The authors also acknowledge experimental assistance by C. Zhan, J. Tian, S. Chodankar and Y. Zhang as well as helpful discussions with F. Lundell, J. Sellberg, M. Nordenström and L. Wågberg.

The experiments were performed at the CHX beamline (11-ID) and the LIX beamline (16-ID) of the National Synchrotron Light Source II, a U.S. Department of Energy (DOE) Office of Science User Facility operated for the DOE Office of Science by Brookhaven National Laboratory under Contract No. DE-SC0012704.

The Center for BioMolecular Structure (CBMS) is primarily supported by the National Institutes of Health, National Institute of General Medical Sciences (NIGMS) through a Center Core P30 Grant (P30GM133893), and by the DOE Office of Biological and Environmental Research (KP1607011).

\end{acknowledgement}

\begin{suppinfo}
{\color{black}
The Supplementary Material includes: a movie showing a 3D rendering of the experimental conditions in the experiment (available at \url{https://play.kth.se/media/SI_XPCS_paper_conditions/0_f8cgf96f}), rheology of the CNF dispersions, a theoretical assessment of pore sizes in random fiber networks, assessment of various fitting methods of the XPCS data, assessment of radiation damage, assessment of gravity-induced sedimentation, sizes and dynamics of GNPs in pure solvent, and results from numerical simulations for other cell sizes.}

\end{suppinfo}

\bibliography{MainManuscipt_XPCS-paper} 

\providecommand{\latin}[1]{#1}
\makeatletter
\providecommand{\doi}
  {\begingroup\let\do\@makeother\dospecials
  \catcode`\{=1 \catcode`\}=2 \doi@aux}
\providecommand{\doi@aux}[1]{\endgroup\texttt{#1}}
\makeatother
\providecommand*\mcitethebibliography{\thebibliography}
\csname @ifundefined\endcsname{endmcitethebibliography}
  {\let\endmcitethebibliography\endthebibliography}{}
\begin{mcitethebibliography}{54}
\providecommand*\natexlab[1]{#1}
\providecommand*\mciteSetBstSublistMode[1]{}
\providecommand*\mciteSetBstMaxWidthForm[2]{}
\providecommand*\mciteBstWouldAddEndPuncttrue
  {\def\EndOfBibitem{\unskip.}}
\providecommand*\mciteBstWouldAddEndPunctfalse
  {\let\EndOfBibitem\relax}
\providecommand*\mciteSetBstMidEndSepPunct[3]{}
\providecommand*\mciteSetBstSublistLabelBeginEnd[3]{}
\providecommand*\EndOfBibitem{}
\mciteSetBstSublistMode{f}
\mciteSetBstMaxWidthForm{subitem}{(\alph{mcitesubitemcount})}
\mciteSetBstSublistLabelBeginEnd
  {\mcitemaxwidthsubitemform\space}
  {\relax}
  {\relax}

\bibitem[Gardel \latin{et~al.}(2004)Gardel, Shin, MacKintosh, Mahadevan,
  Matsudaira, and Weitz]{gardel2004elastic}
Gardel,~M.; Shin,~J.~H.; MacKintosh,~F.; Mahadevan,~L.; Matsudaira,~P.;
  Weitz,~D.~A. Elastic behavior of cross-linked and bundled actin networks.
  \emph{Science} \textbf{2004}, \emph{304}, 1301--1305\relax
\mciteBstWouldAddEndPuncttrue
\mciteSetBstMidEndSepPunct{\mcitedefaultmidpunct}
{\mcitedefaultendpunct}{\mcitedefaultseppunct}\relax
\EndOfBibitem
\bibitem[Gitai(2005)]{gitai2005new}
Gitai,~Z. The new bacterial cell biology: moving parts and subcellular
  architecture. \emph{Cell} \textbf{2005}, \emph{120}, 577--586\relax
\mciteBstWouldAddEndPuncttrue
\mciteSetBstMidEndSepPunct{\mcitedefaultmidpunct}
{\mcitedefaultendpunct}{\mcitedefaultseppunct}\relax
\EndOfBibitem
\bibitem[Jansen \latin{et~al.}(2018)Jansen, Licup, Sharma, Rens, MacKintosh,
  and Koenderink]{jansen2018role}
Jansen,~K.~A.; Licup,~A.~J.; Sharma,~A.; Rens,~R.; MacKintosh,~F.~C.;
  Koenderink,~G.~H. The role of network architecture in collagen mechanics.
  \emph{Biophysical journal} \textbf{2018}, \emph{114}, 2665--2678\relax
\mciteBstWouldAddEndPuncttrue
\mciteSetBstMidEndSepPunct{\mcitedefaultmidpunct}
{\mcitedefaultendpunct}{\mcitedefaultseppunct}\relax
\EndOfBibitem
\bibitem[Ling \latin{et~al.}(2018)Ling, Kaplan, and
  Buehler]{ling2018nanofibrils}
Ling,~S.; Kaplan,~D.~L.; Buehler,~M.~J. Nanofibrils in nature and materials
  engineering. \emph{Nature Reviews Materials} \textbf{2018}, \emph{3},
  1--15\relax
\mciteBstWouldAddEndPuncttrue
\mciteSetBstMidEndSepPunct{\mcitedefaultmidpunct}
{\mcitedefaultendpunct}{\mcitedefaultseppunct}\relax
\EndOfBibitem
\bibitem[Wegst \latin{et~al.}(2015)Wegst, Bai, Saiz, Tomsia, and
  Ritchie]{wegst2015bioinspired}
Wegst,~U.~G.; Bai,~H.; Saiz,~E.; Tomsia,~A.~P.; Ritchie,~R.~O. Bioinspired
  structural materials. \emph{Nature materials} \textbf{2015}, \emph{14},
  23--36\relax
\mciteBstWouldAddEndPuncttrue
\mciteSetBstMidEndSepPunct{\mcitedefaultmidpunct}
{\mcitedefaultendpunct}{\mcitedefaultseppunct}\relax
\EndOfBibitem
\bibitem[Klemm \latin{et~al.}(2011)Klemm, Kramer, Moritz, Lindstr{\"o}m,
  Ankerfors, Gray, and Dorris]{klemm2011nanocelluloses}
Klemm,~D.; Kramer,~F.; Moritz,~S.; Lindstr{\"o}m,~T.; Ankerfors,~M.; Gray,~D.;
  Dorris,~A. Nanocelluloses: a new family of nature-based materials.
  \emph{Angewandte Chemie International Edition} \textbf{2011}, \emph{50},
  5438--5466\relax
\mciteBstWouldAddEndPuncttrue
\mciteSetBstMidEndSepPunct{\mcitedefaultmidpunct}
{\mcitedefaultendpunct}{\mcitedefaultseppunct}\relax
\EndOfBibitem
\bibitem[Li \latin{et~al.}(2021)Li, Chen, Brozena, Zhu, Xu, Driemeier, Dai,
  Rojas, Isogai, W{\aa}gberg, \latin{et~al.} others]{li2021developing}
Li,~T.; Chen,~C.; Brozena,~A.~H.; Zhu,~J.; Xu,~L.; Driemeier,~C.; Dai,~J.;
  Rojas,~O.~J.; Isogai,~A.; W{\aa}gberg,~L., \latin{et~al.}  Developing
  fibrillated cellulose as a sustainable technological material. \emph{Nature}
  \textbf{2021}, \emph{590}, 47--56\relax
\mciteBstWouldAddEndPuncttrue
\mciteSetBstMidEndSepPunct{\mcitedefaultmidpunct}
{\mcitedefaultendpunct}{\mcitedefaultseppunct}\relax
\EndOfBibitem
\bibitem[Ros{\'e}n \latin{et~al.}(2020)Ros{\'e}n, Hsiao, and
  S{\"o}derberg]{rosen2020elucidating}
Ros{\'e}n,~T.; Hsiao,~B.~S.; S{\"o}derberg,~L.~D. Elucidating the Opportunities
  and Challenges for Nanocellulose Spinning. \emph{Advanced Materials}
  \textbf{2020}, 2001238\relax
\mciteBstWouldAddEndPuncttrue
\mciteSetBstMidEndSepPunct{\mcitedefaultmidpunct}
{\mcitedefaultendpunct}{\mcitedefaultseppunct}\relax
\EndOfBibitem
\bibitem[H{\aa}kansson \latin{et~al.}(2014)H{\aa}kansson, Fall, Lundell, Yu,
  Krywka, Roth, Santoro, Kvick, Wittberg, W{\aa}gberg, \latin{et~al.}
  others]{haakansson2014hydrodynamic}
H{\aa}kansson,~K.~M.; Fall,~A.~B.; Lundell,~F.; Yu,~S.; Krywka,~C.;
  Roth,~S.~V.; Santoro,~G.; Kvick,~M.; Wittberg,~L.~P.; W{\aa}gberg,~L.,
  \latin{et~al.}  Hydrodynamic alignment and assembly of nanofibrils resulting
  in strong cellulose filaments. \emph{Nature communications} \textbf{2014},
  \emph{5}, 1--10\relax
\mciteBstWouldAddEndPuncttrue
\mciteSetBstMidEndSepPunct{\mcitedefaultmidpunct}
{\mcitedefaultendpunct}{\mcitedefaultseppunct}\relax
\EndOfBibitem
\bibitem[Mittal \latin{et~al.}(2018)Mittal, Ansari, Gowda.~V, Brouzet, Chen,
  Larsson, Roth, Lundell, Wagberg, Kotov, \latin{et~al.}
  others]{mittal2018multiscale}
Mittal,~N.; Ansari,~F.; Gowda.~V,~K.; Brouzet,~C.; Chen,~P.; Larsson,~P.~T.;
  Roth,~S.~V.; Lundell,~F.; Wagberg,~L.; Kotov,~N.~A., \latin{et~al.}
  Multiscale control of nanocellulose assembly: transferring remarkable
  nanoscale fibril mechanics to macroscale fibers. \emph{ACS nano}
  \textbf{2018}, \emph{12}, 6378--6388\relax
\mciteBstWouldAddEndPuncttrue
\mciteSetBstMidEndSepPunct{\mcitedefaultmidpunct}
{\mcitedefaultendpunct}{\mcitedefaultseppunct}\relax
\EndOfBibitem
\bibitem[Walther \latin{et~al.}(2011)Walther, Timonen, D{\'\i}ez, Laukkanen,
  and Ikkala]{walther2011multifunctional}
Walther,~A.; Timonen,~J.~V.; D{\'\i}ez,~I.; Laukkanen,~A.; Ikkala,~O.
  Multifunctional high-performance biofibers based on wet-extrusion of
  renewable native cellulose nanofibrils. \emph{Advanced Materials}
  \textbf{2011}, \emph{23}, 2924--2928\relax
\mciteBstWouldAddEndPuncttrue
\mciteSetBstMidEndSepPunct{\mcitedefaultmidpunct}
{\mcitedefaultendpunct}{\mcitedefaultseppunct}\relax
\EndOfBibitem
\bibitem[Voisin \latin{et~al.}(2017)Voisin, Bergstr{\"o}m, Liu, and
  Mathew]{voisin2017nanocellulose}
Voisin,~H.; Bergstr{\"o}m,~L.; Liu,~P.; Mathew,~A.~P. Nanocellulose-based
  materials for water purification. \emph{Nanomaterials} \textbf{2017},
  \emph{7}, 57\relax
\mciteBstWouldAddEndPuncttrue
\mciteSetBstMidEndSepPunct{\mcitedefaultmidpunct}
{\mcitedefaultendpunct}{\mcitedefaultseppunct}\relax
\EndOfBibitem
\bibitem[Sharma \latin{et~al.}(2020)Sharma, Sharma, Lindstr{\"o}m, and
  Hsiao]{sharma2020water}
Sharma,~P.~R.; Sharma,~S.~K.; Lindstr{\"o}m,~T.; Hsiao,~B.~S. Water
  Purification: Nanocellulose-Enabled Membranes for Water Purification:
  Perspectives (Adv. Sustainable Syst. 5/2020). \emph{Advanced Sustainable
  Systems} \textbf{2020}, \emph{4}, 2070009\relax
\mciteBstWouldAddEndPuncttrue
\mciteSetBstMidEndSepPunct{\mcitedefaultmidpunct}
{\mcitedefaultendpunct}{\mcitedefaultseppunct}\relax
\EndOfBibitem
\bibitem[Rostami \latin{et~al.}(2021)Rostami, Gordeyeva, Benselfelt,
  Lahchaichi, Hall, Riazanova, Larsson, Ciftci, and
  W{\aa}gberg]{rostami2021hierarchical}
Rostami,~J.; Gordeyeva,~K.; Benselfelt,~T.; Lahchaichi,~E.; Hall,~S.~A.;
  Riazanova,~A.~V.; Larsson,~P.~A.; Ciftci,~G.~C.; W{\aa}gberg,~L. Hierarchical
  build-up of bio-based nanofibrous materials with tunable metal--organic
  framework biofunctionality. \emph{Materials Today} \textbf{2021}, \relax
\mciteBstWouldAddEndPunctfalse
\mciteSetBstMidEndSepPunct{\mcitedefaultmidpunct}
{}{\mcitedefaultseppunct}\relax
\EndOfBibitem
\bibitem[Czaja \latin{et~al.}(2006)Czaja, Krystynowicz, Bielecki, and
  Brown~Jr]{czaja2006microbial}
Czaja,~W.; Krystynowicz,~A.; Bielecki,~S.; Brown~Jr,~R.~M. Microbial
  cellulose—the natural power to heal wounds. \emph{Biomaterials}
  \textbf{2006}, \emph{27}, 145--151\relax
\mciteBstWouldAddEndPuncttrue
\mciteSetBstMidEndSepPunct{\mcitedefaultmidpunct}
{\mcitedefaultendpunct}{\mcitedefaultseppunct}\relax
\EndOfBibitem
\bibitem[Hickey and Pelling(2019)Hickey, and Pelling]{hickey2019cellulose}
Hickey,~R.~J.; Pelling,~A.~E. Cellulose biomaterials for tissue engineering.
  \emph{Frontiers in bioengineering and biotechnology} \textbf{2019}, \emph{7},
  45\relax
\mciteBstWouldAddEndPuncttrue
\mciteSetBstMidEndSepPunct{\mcitedefaultmidpunct}
{\mcitedefaultendpunct}{\mcitedefaultseppunct}\relax
\EndOfBibitem
\bibitem[Eskilson \latin{et~al.}(2020)Eskilson, Lindstr{\"o}m, Sepulveda,
  Shahjamali, G{\"u}ell-Grau, Sivl{\'e}r, Skog, Aronsson, Bj{\"o}rk, Nyberg,
  \latin{et~al.} others]{eskilson2020self}
Eskilson,~O.; Lindstr{\"o}m,~S.~B.; Sepulveda,~B.; Shahjamali,~M.~M.;
  G{\"u}ell-Grau,~P.; Sivl{\'e}r,~P.; Skog,~M.; Aronsson,~C.; Bj{\"o}rk,~E.~M.;
  Nyberg,~N., \latin{et~al.}  Self-Assembly of Mechanoplasmonic Bacterial
  Cellulose--Metal Nanoparticle Composites. \emph{Advanced Functional
  Materials} \textbf{2020}, \emph{30}, 2004766\relax
\mciteBstWouldAddEndPuncttrue
\mciteSetBstMidEndSepPunct{\mcitedefaultmidpunct}
{\mcitedefaultendpunct}{\mcitedefaultseppunct}\relax
\EndOfBibitem
\bibitem[Yao \latin{et~al.}(2018)Yao, Ji, Wang, Wang, and Chen]{yao2018color}
Yao,~J.; Ji,~P.; Wang,~B.; Wang,~H.; Chen,~S. Color-tunable luminescent
  macrofibers based on CdTe QDs-loaded bacterial cellulose nanofibers for pH
  and glucose sensing. \emph{Sensors and Actuators B: Chemical} \textbf{2018},
  \emph{254}, 110--119\relax
\mciteBstWouldAddEndPuncttrue
\mciteSetBstMidEndSepPunct{\mcitedefaultmidpunct}
{\mcitedefaultendpunct}{\mcitedefaultseppunct}\relax
\EndOfBibitem
\bibitem[Hamedi \latin{et~al.}(2014)Hamedi, Hajian, Fall, Hakansson, Salajkova,
  Lundell, Wagberg, and Berglund]{hamedi2014highly}
Hamedi,~M.~M.; Hajian,~A.; Fall,~A.~B.; Hakansson,~K.; Salajkova,~M.;
  Lundell,~F.; Wagberg,~L.; Berglund,~L.~A. Highly conducting, strong
  nanocomposites based on nanocellulose-assisted aqueous dispersions of
  single-wall carbon nanotubes. \emph{ACS nano} \textbf{2014}, \emph{8},
  2467--2476\relax
\mciteBstWouldAddEndPuncttrue
\mciteSetBstMidEndSepPunct{\mcitedefaultmidpunct}
{\mcitedefaultendpunct}{\mcitedefaultseppunct}\relax
\EndOfBibitem
\bibitem[Amiralian \latin{et~al.}(2020)Amiralian, Mustapic, Hossain, Wang,
  Konarova, Tang, Na, Khan, and Rowan]{amiralian2020magnetic}
Amiralian,~N.; Mustapic,~M.; Hossain,~M. S.~A.; Wang,~C.; Konarova,~M.;
  Tang,~J.; Na,~J.; Khan,~A.; Rowan,~A. Magnetic nanocellulose: A potential
  material for removal of dye from water. \emph{Journal of hazardous materials}
  \textbf{2020}, \emph{394}, 122571\relax
\mciteBstWouldAddEndPuncttrue
\mciteSetBstMidEndSepPunct{\mcitedefaultmidpunct}
{\mcitedefaultendpunct}{\mcitedefaultseppunct}\relax
\EndOfBibitem
\bibitem[Wu \latin{et~al.}(2012)Wu, Saito, Fujisawa, Fukuzumi, and
  Isogai]{wu2012ultrastrong}
Wu,~C.-N.; Saito,~T.; Fujisawa,~S.; Fukuzumi,~H.; Isogai,~A. Ultrastrong and
  high gas-barrier nanocellulose/clay-layered composites.
  \emph{Biomacromolecules} \textbf{2012}, \emph{13}, 1927--1932\relax
\mciteBstWouldAddEndPuncttrue
\mciteSetBstMidEndSepPunct{\mcitedefaultmidpunct}
{\mcitedefaultendpunct}{\mcitedefaultseppunct}\relax
\EndOfBibitem
\bibitem[Liu \latin{et~al.}(2011)Liu, Walther, Ikkala, Belova, and
  Berglund]{liu2011clay}
Liu,~A.; Walther,~A.; Ikkala,~O.; Belova,~L.; Berglund,~L.~A. Clay nanopaper
  with tough cellulose nanofiber matrix for fire retardancy and gas barrier
  functions. \emph{Biomacromolecules} \textbf{2011}, \emph{12}, 633--641\relax
\mciteBstWouldAddEndPuncttrue
\mciteSetBstMidEndSepPunct{\mcitedefaultmidpunct}
{\mcitedefaultendpunct}{\mcitedefaultseppunct}\relax
\EndOfBibitem
\bibitem[An \latin{et~al.}(2018)An, Heo, Ji, Bien, and Park]{an2018transparent}
An,~B.~W.; Heo,~S.; Ji,~S.; Bien,~F.; Park,~J.-U. Transparent and flexible
  fingerprint sensor array with multiplexed detection of tactile pressure and
  skin temperature. \emph{Nature communications} \textbf{2018}, \emph{9},
  1--10\relax
\mciteBstWouldAddEndPuncttrue
\mciteSetBstMidEndSepPunct{\mcitedefaultmidpunct}
{\mcitedefaultendpunct}{\mcitedefaultseppunct}\relax
\EndOfBibitem
\bibitem[Okita \latin{et~al.}(2011)Okita, Fujisawa, Saito, and
  Isogai]{okita2011tempo}
Okita,~Y.; Fujisawa,~S.; Saito,~T.; Isogai,~A. TEMPO-oxidized cellulose
  nanofibrils dispersed in organic solvents. \emph{Biomacromolecules}
  \textbf{2011}, \emph{12}, 518--522\relax
\mciteBstWouldAddEndPuncttrue
\mciteSetBstMidEndSepPunct{\mcitedefaultmidpunct}
{\mcitedefaultendpunct}{\mcitedefaultseppunct}\relax
\EndOfBibitem
\bibitem[Wang \latin{et~al.}(2019)Wang, Rosen, Zhan, Chodankar, Chen, Sharma,
  Sharma, Liu, and Hsiao]{wang2019morphology}
Wang,~R.; Rosen,~T.; Zhan,~C.; Chodankar,~S.; Chen,~J.; Sharma,~P.~R.;
  Sharma,~S.~K.; Liu,~T.; Hsiao,~B.~S. Morphology and flow behavior of
  cellulose nanofibers dispersed in glycols. \emph{Macromolecules}
  \textbf{2019}, \emph{52}, 5499--5509\relax
\mciteBstWouldAddEndPuncttrue
\mciteSetBstMidEndSepPunct{\mcitedefaultmidpunct}
{\mcitedefaultendpunct}{\mcitedefaultseppunct}\relax
\EndOfBibitem
\bibitem[Ros{\'e}n \latin{et~al.}(2020)Ros{\'e}n, He, Wang, Zhan, Chodankar,
  Fall, Aulin, Larsson, Lindstr{\"o}m, and Hsiao]{rosen2020cross}
Ros{\'e}n,~T.; He,~H.; Wang,~R.; Zhan,~C.; Chodankar,~S.; Fall,~A.; Aulin,~C.;
  Larsson,~P.~T.; Lindstr{\"o}m,~T.; Hsiao,~B.~S. Cross-Sections of
  Nanocellulose from Wood Analyzed by Quantized Polydispersity of Elementary
  Microfibrils. \emph{ACS nano} \textbf{2020}, \emph{14}, 16743--16754\relax
\mciteBstWouldAddEndPuncttrue
\mciteSetBstMidEndSepPunct{\mcitedefaultmidpunct}
{\mcitedefaultendpunct}{\mcitedefaultseppunct}\relax
\EndOfBibitem
\bibitem[Geng \latin{et~al.}(2017)Geng, Peng, Zhan, Naderi, Sharma, Mao, and
  Hsiao]{geng2017structure}
Geng,~L.; Peng,~X.; Zhan,~C.; Naderi,~A.; Sharma,~P.~R.; Mao,~Y.; Hsiao,~B.~S.
  Structure characterization of cellulose nanofiber hydrogel as functions of
  concentration and ionic strength. \emph{Cellulose} \textbf{2017}, \emph{24},
  5417--5429\relax
\mciteBstWouldAddEndPuncttrue
\mciteSetBstMidEndSepPunct{\mcitedefaultmidpunct}
{\mcitedefaultendpunct}{\mcitedefaultseppunct}\relax
\EndOfBibitem
\bibitem[Ros{\'e}n \latin{et~al.}(2021)Ros{\'e}n, Wang, He, Zhan, Chodankar,
  and Hsiao]{rosen2021understanding}
Ros{\'e}n,~T.; Wang,~R.; He,~H.; Zhan,~C.; Chodankar,~S.; Hsiao,~B.~S.
  Understanding Ion-Induced Assembly of Cellulose Nanofibrillar Gels through
  Shear-Free Mixing and In Situ Scanning-SAXS. \emph{Nanoscale advances}
  \textbf{2021}, \emph{3}, 4940--4951\relax
\mciteBstWouldAddEndPuncttrue
\mciteSetBstMidEndSepPunct{\mcitedefaultmidpunct}
{\mcitedefaultendpunct}{\mcitedefaultseppunct}\relax
\EndOfBibitem
\bibitem[Geng \latin{et~al.}(2018)Geng, Mittal, Zhan, Ansari, Sharma, Peng,
  Hsiao, and S{\"o}derberg]{geng2018understanding}
Geng,~L.; Mittal,~N.; Zhan,~C.; Ansari,~F.; Sharma,~P.~R.; Peng,~X.;
  Hsiao,~B.~S.; S{\"o}derberg,~L.~D. Understanding the mechanistic behavior of
  highly charged cellulose nanofibers in aqueous systems. \emph{Macromolecules}
  \textbf{2018}, \emph{51}, 1498--1506\relax
\mciteBstWouldAddEndPuncttrue
\mciteSetBstMidEndSepPunct{\mcitedefaultmidpunct}
{\mcitedefaultendpunct}{\mcitedefaultseppunct}\relax
\EndOfBibitem
\bibitem[Ros{\'e}n \latin{et~al.}(2021)Ros{\'e}n, Wang, He, Zhan, Chodankar,
  and Hsiao]{rosen2021shear}
Ros{\'e}n,~T.; Wang,~R.; He,~H.; Zhan,~C.; Chodankar,~S.; Hsiao,~B.~S.
  Shear-Free Mixing to Achieve Accurate Temporospatial Nanoscale Kinetics
  through Scanning-SAXS: Ion-Induced Phase Transition of Dispersed Cellulose
  Nanocrystals. \emph{Lab on a Chip} \textbf{2021}, \emph{21}, 1084--1095\relax
\mciteBstWouldAddEndPuncttrue
\mciteSetBstMidEndSepPunct{\mcitedefaultmidpunct}
{\mcitedefaultendpunct}{\mcitedefaultseppunct}\relax
\EndOfBibitem
\bibitem[Ogston(1958)]{ogston1958spaces}
Ogston,~A. The spaces in a uniform random suspension of fibres.
  \emph{Transactions of the Faraday Society} \textbf{1958}, \emph{54},
  1754--1757\relax
\mciteBstWouldAddEndPuncttrue
\mciteSetBstMidEndSepPunct{\mcitedefaultmidpunct}
{\mcitedefaultendpunct}{\mcitedefaultseppunct}\relax
\EndOfBibitem
\bibitem[Chatterjee(2012)]{chatterjee2012simple}
Chatterjee,~A.~P. A simple model for the pore size distribution in random fibre
  networks. \emph{Journal of Physics: Condensed Matter} \textbf{2012},
  \emph{24}, 375106\relax
\mciteBstWouldAddEndPuncttrue
\mciteSetBstMidEndSepPunct{\mcitedefaultmidpunct}
{\mcitedefaultendpunct}{\mcitedefaultseppunct}\relax
\EndOfBibitem
\bibitem[Madsen \latin{et~al.}(2010)Madsen, Leheny, Guo, Sprung, and
  Czakkel]{madsen2010beyond}
Madsen,~A.; Leheny,~R.~L.; Guo,~H.; Sprung,~M.; Czakkel,~O. Beyond simple
  exponential correlation functions and equilibrium dynamics in x-ray photon
  correlation spectroscopy. \emph{New Journal of Physics} \textbf{2010},
  \emph{12}, 055001\relax
\mciteBstWouldAddEndPuncttrue
\mciteSetBstMidEndSepPunct{\mcitedefaultmidpunct}
{\mcitedefaultendpunct}{\mcitedefaultseppunct}\relax
\EndOfBibitem
\bibitem[Provencher(1982)]{provencher1982contin}
Provencher,~S.~W. CONTIN: a general purpose constrained regularization program
  for inverting noisy linear algebraic and integral equations. \emph{Computer
  Physics Communications} \textbf{1982}, \emph{27}, 229--242\relax
\mciteBstWouldAddEndPuncttrue
\mciteSetBstMidEndSepPunct{\mcitedefaultmidpunct}
{\mcitedefaultendpunct}{\mcitedefaultseppunct}\relax
\EndOfBibitem
\bibitem[Andrews \latin{et~al.}(2018)Andrews, Narayanan, Zhang, Kuzmenko, and
  Ilavsky]{andrews2018inverse}
Andrews,~R.~N.; Narayanan,~S.; Zhang,~F.; Kuzmenko,~I.; Ilavsky,~J. Inverse
  transformation: unleashing spatially heterogeneous dynamics with an
  alternative approach to XPCS data analysis. \emph{Journal of applied
  crystallography} \textbf{2018}, \emph{51}, 35--46\relax
\mciteBstWouldAddEndPuncttrue
\mciteSetBstMidEndSepPunct{\mcitedefaultmidpunct}
{\mcitedefaultendpunct}{\mcitedefaultseppunct}\relax
\EndOfBibitem
\bibitem[Koppel(1972)]{koppel1972analysis}
Koppel,~D.~E. Analysis of Macromolecular Polydispersity in Intensity
  Correlation Spectroscopy: The Method of Cumulants. \emph{The Journal of
  Chemical Physics} \textbf{1972}, \emph{57}, 4814--4820\relax
\mciteBstWouldAddEndPuncttrue
\mciteSetBstMidEndSepPunct{\mcitedefaultmidpunct}
{\mcitedefaultendpunct}{\mcitedefaultseppunct}\relax
\EndOfBibitem
\bibitem[Tracy and Pecora(1992)Tracy, and Pecora]{tracy1992synthesis}
Tracy,~M.~A.; Pecora,~R. Synthesis, characterization, and dynamics of a
  rod/sphere composite liquid. \emph{Macromolecules} \textbf{1992}, \emph{25},
  337--349\relax
\mciteBstWouldAddEndPuncttrue
\mciteSetBstMidEndSepPunct{\mcitedefaultmidpunct}
{\mcitedefaultendpunct}{\mcitedefaultseppunct}\relax
\EndOfBibitem
\bibitem[Kluijtmans \latin{et~al.}(2000)Kluijtmans, Koenderink, and
  Philipse]{kluijtmans2000self}
Kluijtmans,~S.~G.; Koenderink,~G.~H.; Philipse,~A.~P. Self-diffusion and
  sedimentation of tracer spheres in (semi) dilute dispersions of rigid
  colloidal rods. \emph{Physical Review E} \textbf{2000}, \emph{61}, 626\relax
\mciteBstWouldAddEndPuncttrue
\mciteSetBstMidEndSepPunct{\mcitedefaultmidpunct}
{\mcitedefaultendpunct}{\mcitedefaultseppunct}\relax
\EndOfBibitem
\bibitem[Pryamitsyn and Ganesan(2008)Pryamitsyn, and
  Ganesan]{pryamitsyn2008dynamics}
Pryamitsyn,~V.; Ganesan,~V. Dynamics of probe diffusion in rod solutions.
  \emph{Physical review letters} \textbf{2008}, \emph{100}, 128302\relax
\mciteBstWouldAddEndPuncttrue
\mciteSetBstMidEndSepPunct{\mcitedefaultmidpunct}
{\mcitedefaultendpunct}{\mcitedefaultseppunct}\relax
\EndOfBibitem
\bibitem[Metzler \latin{et~al.}(2014)Metzler, Jeon, Cherstvy, and
  Barkai]{metzler2014anomalous}
Metzler,~R.; Jeon,~J.-H.; Cherstvy,~A.~G.; Barkai,~E. Anomalous diffusion
  models and their properties: non-stationarity, non-ergodicity, and ageing at
  the centenary of single particle tracking. \emph{Physical Chemistry Chemical
  Physics} \textbf{2014}, \emph{16}, 24128--24164\relax
\mciteBstWouldAddEndPuncttrue
\mciteSetBstMidEndSepPunct{\mcitedefaultmidpunct}
{\mcitedefaultendpunct}{\mcitedefaultseppunct}\relax
\EndOfBibitem
\bibitem[Pal \latin{et~al.}(2021)Pal, Kamal, Zinn, Dhont, and
  Schurtenberger]{pal2021anisotropic}
Pal,~A.; Kamal,~M.~A.; Zinn,~T.; Dhont,~J.~K.; Schurtenberger,~P. Anisotropic
  dynamics of magnetic colloidal cubes studied by x-ray photon correlation
  spectroscopy. \emph{Physical Review Materials} \textbf{2021}, \emph{5},
  035603\relax
\mciteBstWouldAddEndPuncttrue
\mciteSetBstMidEndSepPunct{\mcitedefaultmidpunct}
{\mcitedefaultendpunct}{\mcitedefaultseppunct}\relax
\EndOfBibitem
\bibitem[Reiser \latin{et~al.}(2020)Reiser, Hallmann, M{\"o}ller, Kazarian,
  Orsi, Randolph, Rahmann, Westermeier, Stellamanns, Sprung, \latin{et~al.}
  others]{reiser2020nanoscale}
Reiser,~M.; Hallmann,~J.; M{\"o}ller,~J.; Kazarian,~K.; Orsi,~D.; Randolph,~L.;
  Rahmann,~H.; Westermeier,~F.; Stellamanns,~E.; Sprung,~M., \latin{et~al.}
  Nanoscale Rigidity in Cross-Linked Micelle Networks Revealed by XPCS
  Nanorheology. \emph{arXiv preprint arXiv:2010.08267} \textbf{2020}, \relax
\mciteBstWouldAddEndPunctfalse
\mciteSetBstMidEndSepPunct{\mcitedefaultmidpunct}
{}{\mcitedefaultseppunct}\relax
\EndOfBibitem
\bibitem[Kwa{\'s}niewski \latin{et~al.}(2014)Kwa{\'s}niewski, Fluerasu, and
  Madsen]{kwasniewski2014anomalous}
Kwa{\'s}niewski,~P.; Fluerasu,~A.; Madsen,~A. Anomalous dynamics at the
  hard-sphere glass transition. \emph{Soft Matter} \textbf{2014}, \emph{10},
  8698--8704\relax
\mciteBstWouldAddEndPuncttrue
\mciteSetBstMidEndSepPunct{\mcitedefaultmidpunct}
{\mcitedefaultendpunct}{\mcitedefaultseppunct}\relax
\EndOfBibitem
\bibitem[Gowda \latin{et~al.}(2022)Gowda, Ros{\'e}n, Roth, S{\"o}derberg, and
  Lundell]{gowda2022nanofibril}
Gowda,~V.~K.; Ros{\'e}n,~T.; Roth,~S.~V.; S{\"o}derberg,~L.~D.; Lundell,~F.
  Nanofibril Alignment during Assembly Revealed by an X-ray Scattering-Based
  Digital Twin. \emph{ACS nano} \textbf{2022}, \emph{16}, 2120--2132\relax
\mciteBstWouldAddEndPuncttrue
\mciteSetBstMidEndSepPunct{\mcitedefaultmidpunct}
{\mcitedefaultendpunct}{\mcitedefaultseppunct}\relax
\EndOfBibitem
\bibitem[Ernst \latin{et~al.}(2017)Ernst, John, Guenther, Wagner, Schaefer, and
  Lehr]{ernst2017model}
Ernst,~M.; John,~T.; Guenther,~M.; Wagner,~C.; Schaefer,~U.~F.; Lehr,~C.-M. A
  model for the transient subdiffusive behavior of particles in mucus.
  \emph{Biophysical journal} \textbf{2017}, \emph{112}, 172--179\relax
\mciteBstWouldAddEndPuncttrue
\mciteSetBstMidEndSepPunct{\mcitedefaultmidpunct}
{\mcitedefaultendpunct}{\mcitedefaultseppunct}\relax
\EndOfBibitem
\bibitem[Xu \latin{et~al.}(2021)Xu, Dai, Bu, Yang, Zhang, Man, Zhang, Doi, and
  Yan]{xu2021enhanced}
Xu,~Z.; Dai,~X.; Bu,~X.; Yang,~Y.; Zhang,~X.; Man,~X.; Zhang,~X.; Doi,~M.;
  Yan,~L.-T. Enhanced Heterogeneous Diffusion of Nanoparticles in Semiflexible
  Networks. \emph{ACS nano} \textbf{2021}, \emph{15}, 4608--4616\relax
\mciteBstWouldAddEndPuncttrue
\mciteSetBstMidEndSepPunct{\mcitedefaultmidpunct}
{\mcitedefaultendpunct}{\mcitedefaultseppunct}\relax
\EndOfBibitem
\bibitem[Saito \latin{et~al.}(2007)Saito, Kimura, Nishiyama, and
  Isogai]{saito2007cellulose}
Saito,~T.; Kimura,~S.; Nishiyama,~Y.; Isogai,~A. Cellulose nanofibers prepared
  by TEMPO-mediated oxidation of native cellulose. \emph{Biomacromolecules}
  \textbf{2007}, \emph{8}, 2485--2491\relax
\mciteBstWouldAddEndPuncttrue
\mciteSetBstMidEndSepPunct{\mcitedefaultmidpunct}
{\mcitedefaultendpunct}{\mcitedefaultseppunct}\relax
\EndOfBibitem
\bibitem[Katz \latin{et~al.}(1984)Katz, Beatson, \latin{et~al.}
  others]{katz1984determination}
Katz,~S.; Beatson,~R.~P., \latin{et~al.}  The determination of strong and weak
  acidic groups in sulfite pulps. \emph{Svensk papperstidning} \textbf{1984},
  \emph{87}, 48--53\relax
\mciteBstWouldAddEndPuncttrue
\mciteSetBstMidEndSepPunct{\mcitedefaultmidpunct}
{\mcitedefaultendpunct}{\mcitedefaultseppunct}\relax
\EndOfBibitem
\bibitem[Lumma \latin{et~al.}(2000)Lumma, Lurio, Mochrie, and
  Sutton]{lumma2000area}
Lumma,~D.; Lurio,~L.; Mochrie,~S.; Sutton,~M. Area Detector Based Photon
  Correlation in the Regime of Short Data Batches: Data Reduction for Dynamic
  X-ray Scattering. \emph{Review of Scientific Instruments} \textbf{2000},
  \emph{71}, 3274--3289\relax
\mciteBstWouldAddEndPuncttrue
\mciteSetBstMidEndSepPunct{\mcitedefaultmidpunct}
{\mcitedefaultendpunct}{\mcitedefaultseppunct}\relax
\EndOfBibitem
\bibitem[Fluerasu \latin{et~al.}(2007)Fluerasu, Moussa{\"\i}d, Madsen, and
  Schofield]{fluerasu2007slow}
Fluerasu,~A.; Moussa{\"\i}d,~A.; Madsen,~A.; Schofield,~A. Slow dynamics and
  aging in colloidal gels studied by x-ray photon correlation spectroscopy.
  \emph{Physical Review E} \textbf{2007}, \emph{76}, 010401\relax
\mciteBstWouldAddEndPuncttrue
\mciteSetBstMidEndSepPunct{\mcitedefaultmidpunct}
{\mcitedefaultendpunct}{\mcitedefaultseppunct}\relax
\EndOfBibitem
\bibitem[Marino(2007)]{MATLAB_RILT}
Marino,~I.-G. rilt - Regularized Inverse Laplace Transform.
  https://www.mathworks.com/matlabcentral/fileexchange/6523-rilt, 2007; MATLAB
  Central File Exchange. Retrieved June 3, 2021\relax
\mciteBstWouldAddEndPuncttrue
\mciteSetBstMidEndSepPunct{\mcitedefaultmidpunct}
{\mcitedefaultendpunct}{\mcitedefaultseppunct}\relax
\EndOfBibitem
\bibitem[Sagdeev \latin{et~al.}(2017)Sagdeev, Fomina, and
  Abdulagatov]{sagdeev2017density}
Sagdeev,~D.~I.; Fomina,~M.~G.; Abdulagatov,~I.~M. Density and viscosity of
  propylene glycol at high temperatures and high pressures. \emph{Fluid Phase
  Equilibria} \textbf{2017}, \emph{450}, 99--111\relax
\mciteBstWouldAddEndPuncttrue
\mciteSetBstMidEndSepPunct{\mcitedefaultmidpunct}
{\mcitedefaultendpunct}{\mcitedefaultseppunct}\relax
\EndOfBibitem
\bibitem[Feigin \latin{et~al.}(1987)Feigin, Svergun, \latin{et~al.}
  others]{feigin1987structure}
Feigin,~L.; Svergun,~D.~I., \latin{et~al.}  \emph{Structure analysis by
  small-angle X-ray and neutron scattering}; Springer, 1987; Vol.~1\relax
\mciteBstWouldAddEndPuncttrue
\mciteSetBstMidEndSepPunct{\mcitedefaultmidpunct}
{\mcitedefaultendpunct}{\mcitedefaultseppunct}\relax
\EndOfBibitem
\end{mcitethebibliography}


\providecommand{\latin}[1]{#1}
\makeatletter
\providecommand{\doi}
  {\begingroup\let\do\@makeother\dospecials
  \catcode`\{=1 \catcode`\}=2 \doi@aux}
\providecommand{\doi@aux}[1]{\endgroup\texttt{#1}}
\makeatother
\providecommand*\mcitethebibliography{\thebibliography}
\csname @ifundefined\endcsname{endmcitethebibliography}
  {\let\endmcitethebibliography\endthebibliography}{}
\begin{mcitethebibliography}{5}
\providecommand*\natexlab[1]{#1}
\providecommand*\mciteSetBstSublistMode[1]{}
\providecommand*\mciteSetBstMaxWidthForm[2]{}
\providecommand*\mciteBstWouldAddEndPuncttrue
  {\def\EndOfBibitem{\unskip.}}
\providecommand*\mciteBstWouldAddEndPunctfalse
  {\let\EndOfBibitem\relax}
\providecommand*\mciteSetBstMidEndSepPunct[3]{}
\providecommand*\mciteSetBstSublistLabelBeginEnd[3]{}
\providecommand*\EndOfBibitem{}
\mciteSetBstSublistMode{f}
\mciteSetBstMaxWidthForm{subitem}{(\alph{mcitesubitemcount})}
\mciteSetBstSublistLabelBeginEnd
  {\mcitemaxwidthsubitemform\space}
  {\relax}
  {\relax}

\bibitem[Sagdeev \latin{et~al.}(2017)Sagdeev, Fomina, and
  Abdulagatov]{sagdeev2017density}
Sagdeev,~D.~I.; Fomina,~M.~G.; Abdulagatov,~I.~M. Density and viscosity of
  propylene glycol at high temperatures and high pressures. \emph{Fluid Phase
  Equilibria} \textbf{2017}, \emph{450}, 99--111\relax
\mciteBstWouldAddEndPuncttrue
\mciteSetBstMidEndSepPunct{\mcitedefaultmidpunct}
{\mcitedefaultendpunct}{\mcitedefaultseppunct}\relax
\EndOfBibitem
\bibitem[Ogston(1958)]{ogston1958spaces}
Ogston,~A. The spaces in a uniform random suspension of fibres.
  \emph{Transactions of the Faraday Society} \textbf{1958}, \emph{54},
  1754--1757\relax
\mciteBstWouldAddEndPuncttrue
\mciteSetBstMidEndSepPunct{\mcitedefaultmidpunct}
{\mcitedefaultendpunct}{\mcitedefaultseppunct}\relax
\EndOfBibitem
\bibitem[Chatterjee(2012)]{chatterjee2012simple}
Chatterjee,~A.~P. A simple model for the pore size distribution in random fibre
  networks. \emph{Journal of Physics: Condensed Matter} \textbf{2012},
  \emph{24}, 375106\relax
\mciteBstWouldAddEndPuncttrue
\mciteSetBstMidEndSepPunct{\mcitedefaultmidpunct}
{\mcitedefaultendpunct}{\mcitedefaultseppunct}\relax
\EndOfBibitem
\bibitem[Sas()]{SasView}
SasView 5.0.4 documentation: Polydispersity and Orientational Distributions.
  \url{https://www.sasview.org/docs/user/qtgui/Perspectives/Fitting/pd/polydispersity.html},
  Accessed: 2022-04-14\relax
\mciteBstWouldAddEndPuncttrue
\mciteSetBstMidEndSepPunct{\mcitedefaultmidpunct}
{\mcitedefaultendpunct}{\mcitedefaultseppunct}\relax
\EndOfBibitem
\end{mcitethebibliography}

\end{document}



\begin{figure}[t]
\centering
\includegraphics[width=0.99\textwidth]{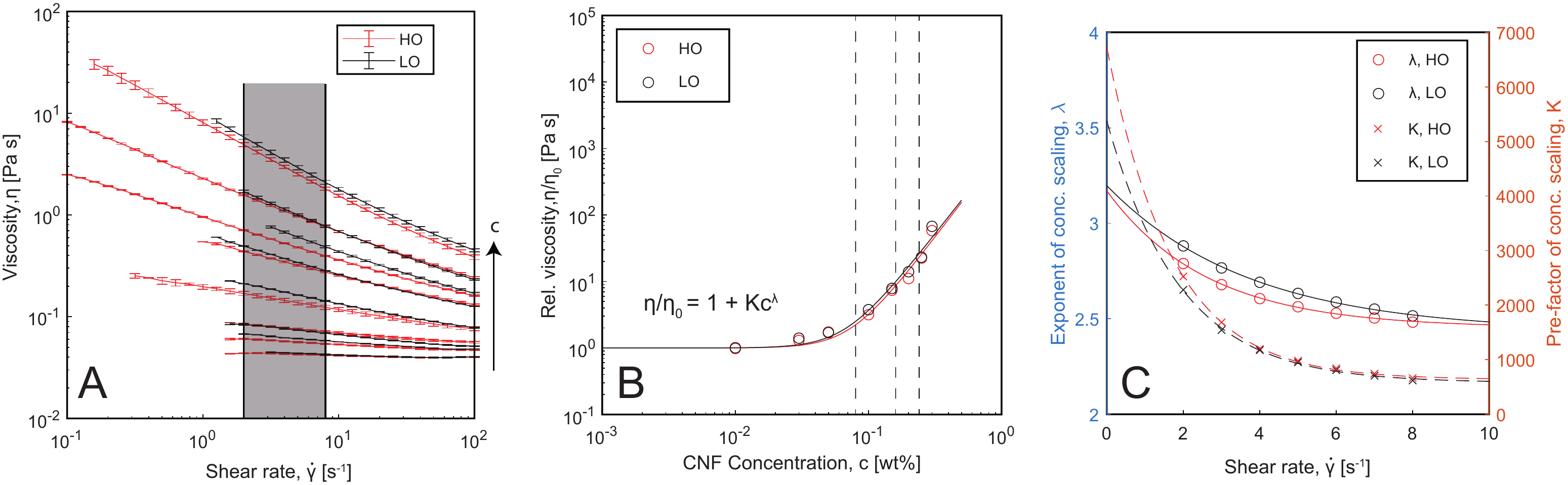}
\caption{Steady shear viscosity measurements of CNF dispersions in the study; red and black curves/symbols indicate high and low degree of oxidation of dispersions (HO and LO), respectively; (A) dynamic viscosity $\eta$ of dispersions with varying shear rates $\dot{\gamma}$ at concentrations $c=0.01, 0.03, 0.05, 0.1, 0.15, 0.2, 0.25$~and~0.3~wt\%; the shaded region shows the set of shear rates used to estimate zero-shear viscosity; (B) relative viscosity $\eta/\eta_0$ at $\dot{\gamma}=5$~s$^{-1}$; the dashed lines indicate the concentrations used in the XPCS-study; (C) the extracted parameters $K$ and $\lambda$ as function of shear rate $\dot{\gamma}$.}
\label{fig:Rheology}
\end{figure}

\section{Rheology of CNF dispersions}
Steady shear viscosity of the two CNF dispersions at high oxidation (HO) and low oxidation (LO) was measured according to the procedure described in the main manuscript. The results are shown in Fig-~\ref{fig:Rheology}A-C. In Fig~\ref{fig:Rheology}A, the viscosity $\eta$ versus shear rate $\dot{\gamma}$ is shown of the two dispersions at 8 different concentrations ($c=0.01, 0.03, 0.05, 0.1, 0.15, 0.2, 0.25$~and~0.3~wt\%), where we can see increasing shear-thinning behavior at increasing concentration. The relative viscosity $\eta/\eta_0$ is determined using the viscosity of $\eta_0 = 0.042$~Pa$\cdot$s for propylene glycol (PG)~\cite{sagdeev2017density} and the concentration dependent data at a fixed shear rate $\dot{\gamma}$ is fitted with a function of form $\eta/\eta_0=1+K(\dot{\gamma})c^{\lambda(\dot{\gamma})}$, illustrated in Fig.~\ref{fig:Rheology}B. The dashed lines in the figure show the three concentrations used in the XPCS-study ($c=0.08, 0.16$~and~$0.24$~wt\%). The shear rate dependence of parameters $K$ and $\lambda$ in region $\dot{\gamma}=2$~to~8~s$^{-1}$ is illustrated in Fig.~\ref{fig:Rheology}C. The data is fitted to exponential functions:\\
\begin{equation}
\lambda(\dot{\gamma})=C_1\exp(-C_2\dot{\gamma})+C_3
\end{equation} 
\begin{equation}
K(\dot{\gamma})=C_4\exp(-C_5\dot{\gamma})+C_6
\end{equation}
~\\
to estimate the zero-shear parameters $\lambda(0)=C_1+C_3$ and $K(0)=C_4+C_6$ ($C_{1-6}$ are arbitrary fitting parameters). The zero-shear viscosity of the two CNF dispersions at various concentrations is then estimated as $\eta/\eta_0=1+K(0)c^{\lambda(0)}$. The expressions to describe the zero-shear viscosity for the two dispersions are:\\
\begin{equation}
\eta^{(HO)}/\eta_0 = 1+5372c^{3.20}
\end{equation} 
\begin{equation}
\eta^{(LO)}/\eta_0 = 1+6722c^{3.17}.
\end{equation} 
Using these expressions, we can also estimate the overlap concentrations $c^*$ where CNF interactions dominate the viscosity scaling through $c^*=K(0)^{-1/\lambda(0)}$. This is found to be $c^*=0.068$ and $0.062$~wt\% for the HO and LO CNF, respectively. This also indicates that the lowest concentration in the experiment of $c=0.08$~wt\% is only slightly above this overlap concentration.

\begin{figure}[t]
\centering
\includegraphics[width=0.4\textwidth]{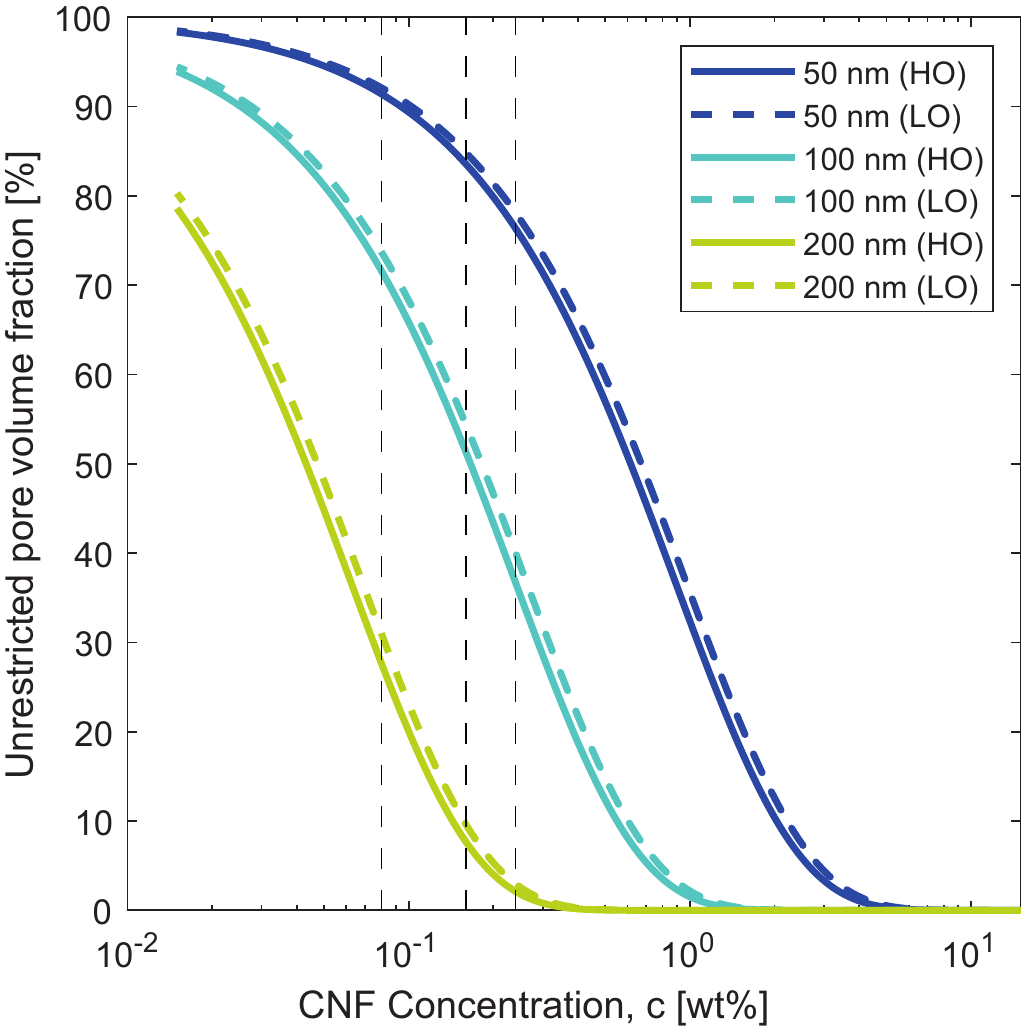}
\caption{Unrestricted pore volume fraction $UPV$ for spherical nanoparticles in typical random fiber networks of stiff cylindrical fibers with the same equivalent diameter as the CNFs in the experiments according to theory by Ogston~\cite{ogston1958spaces}. }
\label{fig:PoreSizeAnalysis}
\end{figure}

\section{Pore sizes in random fiber networks}
The probability of a randomly placed sphere in a fiber network to not be in contact with a fiber is called the unrestricted pore volume fraction ($UPV$) and is calculated through the theory by Ogston~\cite{ogston1958spaces} using expressions by Chatterjee~\cite{chatterjee2012simple}:\\
\begin{equation}
UPV = (1-\Phi)^{(R_\text{sphere}/R_\text{fiber})(R_\text{sphere}/R_\text{fiber}+2)}
\end{equation}
~\\
where $\Phi$ is the volume fraction of fibers. Fig.~\ref{fig:PoreSizeAnalysis} shows the theoretical $UPV$ at various concentrations $c$ of the fiber network assuming the CNFs as long rigid cylinders with radius $R_\text{fiber}$ determined from the mean equivalent cross-sectional size of the CNFs found with SAXS (Fig.~1C in main manuscript). Here, $R_\text{fiber} = 2.1$~and~2.3~nm for HO and LO CNF, respectively. The black vertical dashed lines indicate the concentrations used in the XPCS-study ($c=0.08, 0.16$~and~$0.24$~wt\%). Note that $UPV>90\%$ for 50~nm GNPs at 0.08~wt\%, \emph{i.e.} that GNPs are almost moving freely in the network. In contrast, 200~nm GNPs at 0.24~wt\% are practically always in contact with the network ($UPV<5\%$). 

\begin{figure}[tbp]
\centering
\includegraphics[width=0.99\textwidth]{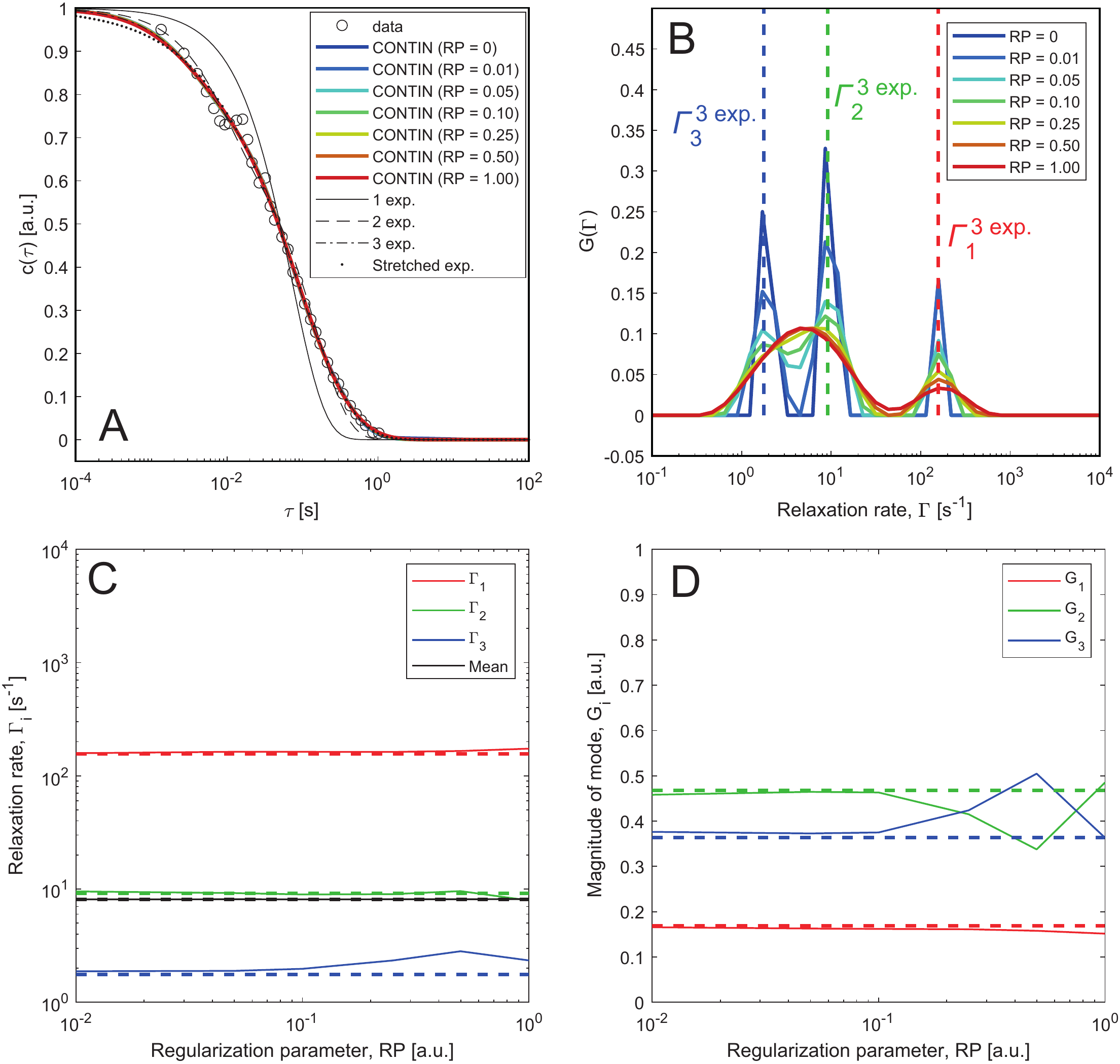}
\caption{Assessment of various fitting methods applied to a dataset of $c(\tau)$ for 200~nm GNPs in 0.24~wt\% HO CNF at $q=0.0025$~nm$^{-1}$; (A) The data $c(\tau)$ with curves of the various fits; (B) relaxation rate distributions $G(\Gamma)$ from the CONTIN fit at various regularization parameters (RP) with dashed lines showing the relaxation rates found by fitting 3 exponentials; (C) and (D) the relaxation rates and magnitude of dynamic modes at various RP, respectively.}
\label{fig:Fitting_noNoise}
\end{figure}

\begin{figure}[tbp]
\centering
\includegraphics[width=0.99\textwidth]{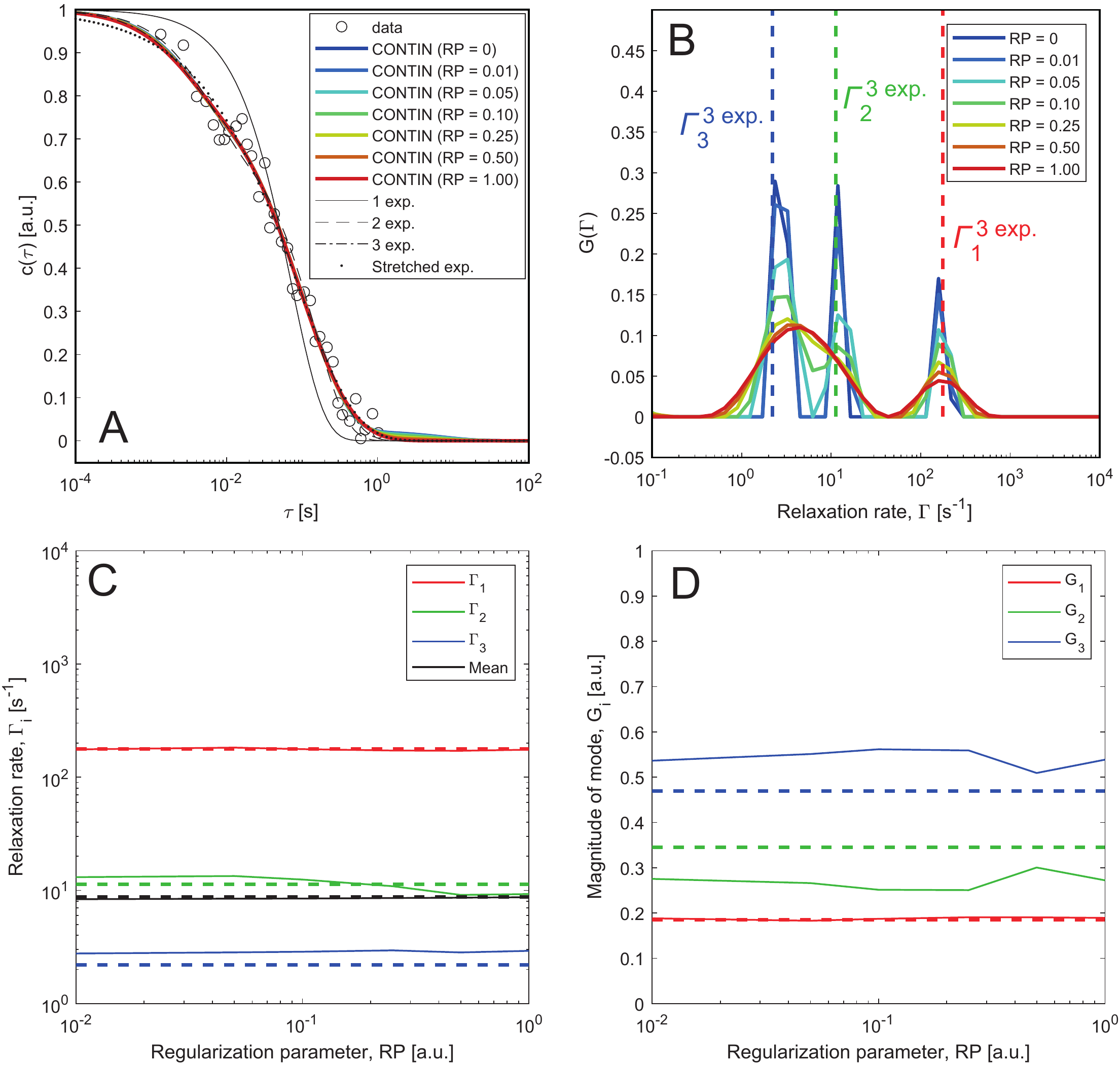}
\caption{The figure shows exactly the same assessment done to the same data as in Fig.~\ref{fig:Fitting_noNoise}, but with added random noise $c_\epsilon\in[-0.05,0.05]$ to the original data.}
\label{fig:Fitting_wNoise}
\end{figure}

\begin{table}[tbp]
\begin{tabular}{l|c|c|c|c|c|c|c|}
         	      & $\bar{\Gamma}$ [s$^{-1}$] & $\Gamma_1$ [s$^{-1}$] & $\Gamma_2$ [s$^{-1}$] & $\Gamma_3$ [s$^{-1}$] & $PDI$ & $\gamma$ & $R^2$ \\
         	      \hline
         	      \hline
Stretched exponential & 3.57 & - & - & - & - & 0.59 & 0.9960   \\
$c=\exp(-2[\bar{\Gamma}\tau]^\gamma)$ & (3.44) & - & - & - & - & (0.57) & (0.9817)\\
\hline
\hline
One exponential & 6.86 & - & - & - & - & - & 0.9054   \\
$c=\exp(-2\bar{\Gamma}\tau)$ & (6.98) & - & - & - & - & - & (0.8692)\\
\hline
Two exponentials & 7.93 & 59.2 & 3.27 & - & 3.20 & - & 0.9909   \\
$c=\sum_{i=1}^2 G_i\exp(-2\Gamma_i\tau)$ & (8.68) & (95.1) & (3.42) & - & (4.00) & - & (0.9787)\\
\hline
Three exponentials & 8.13 & 156.6 & 9.18 & 1.76 & 6.51 & - & 0.9971   \\
$c=\sum_{i=1}^3 G_i\exp(-2\Gamma_i\tau)$ & (8.73) & (177.2) & (11.3) & (2.20) & (6.22) & - & (0.9842)\\
\hline
Four exponentials & 8.10 & -& - & - & 6.59 & - & 0.9971   \\
$c=\sum_{i=1}^4 G_i\exp(-2\Gamma_i\tau)$ & (8.74) & - & - & - & (6.23) & - & (0.9842)\\
\hline
\hline
CONTIN (RP=0) & 8.06 & 158.3 & 9.71 & 1.94 & 6.56 & - & 0.9971 \\
(w. log-normal deconv.) & (8.35) & (173.0) & (11.9) & (2.73) & (6.06) & - & (0.9845)\\
$c=\sum_{i=1}^N G_i\exp(-2\Gamma_i\tau)$ &&&&&&&\\
\hline
CONTIN (RP=0.01) & 8.10 & 158.3 & 9.53 & 1.88 & 6.56 & - & 0.9971   \\
(w. log-normal deconv.) & (8.36) & (175.9) & (13.1) & (2.77) & (6.22) & - & (0.9845)\\
\hline
CONTIN (RP=0.05) & 8.12 & 162.9 & 9.21 & 1.89 & 7.09 & - & 0.9971   \\
(w. log-normal deconv.) & (8.45) & (182.4) & (13.4) & (2.83) & (6.53) & - & (0.9844)\\
\hline
CONTIN (RP=0.1) & 8.11 & 162.8 & 8.95 & 1.97 & 7.16 & - & 0.9970   \\
(w. log-normal deconv.) & (8.46) & (176.8) & (12.5) & (2.87) & (6.48) & - & (0.9842)\\
\hline
CONTIN (RP=0.25) & 8.12 & 162.4 & 9.01 & 2.34 & 7.81 & - & 0.9969  \\
(w. log-normal deconv.) & (8.55) & (171.9) & (10.9) & (2.95) & (6.97) & - & (0.9840)\\
\hline
CONTIN (RP=0.5) & 8.13 & 165.5 & 9.59 & 2.82 & 8.52 & - & 0.9969   \\
(w. log-normal deconv.) & (8.62) & (171.1) & (9.12) & (2.83) & (7.49) & - & (0.9839)\\
\hline
CONTIN (RP=1.0) & 8.15 & 173.5 & 7.93 & 2.34 & 9.90 & - & 0.9967   \\
(w. log-normal deconv.) & (8.70) & (174.8) & (9.27) & (2.92) & (8.38) & - & (0.9837)\\
\hline
\end{tabular}
\caption{Table of parameters from the various types of fitting methods applied to the data of 200~nm GNPs in HO-CNF at 0.24~wt\% and  $q=0.0025$~nm$^{-1}$. Values in parenthesis are results when adding random $\pm 5$~\% noise to the data (see Fig.~\ref{fig:Fitting_wNoise}A).}
\label{tab:Tab1}
\end{table}

\section{Assessment of various fitting methods of XPCS data}
The square first order autocorrelation $c(\tau)$ from one representative XPCS experiment of 200~nm GNPs in HO CNF at 0.24~wt\% ($q=0.0025$~nm$^{-1}$) was used to compare various fitting methods (see Fig.~\ref{fig:Fitting_noNoise}A). Three methods were assessed:~\\
\begin{enumerate}
\item The correlation is assumed the form:\\
\begin{equation}
c(\tau)=\sum_i G^{(d)}_i\exp(-2\Gamma^{(d)}_i\tau), 
\end{equation}
~\\
where the distribution $G^{(d)}_i$ of pre-determined log-spaced relaxation rates $\Gamma^{(d)}_i$ is found using the CONTIN-algorithm with various values of the regularization parameter (RP) using the constraint that $\sum_{i=1}^N G^{(d)}_i=1$. The value determines the level of smoothing of the distribution (see Fig.~\ref{fig:Fitting_noNoise}B). The distribution is used to extract the relaxation rates $\Gamma_{1-3}$ and magnitudes $G_{1-3}$ of the three dynamic modes according to the lognormal deconvolution described in the main manuscript (see Fig.~\ref{fig:Fitting_noNoise}C-D).

\item The correlation is assumed a similar form as the one above:\\
\begin{equation}
c(\tau)=\sum_{i=1}^N G_i\exp(-2\Gamma_i\tau), 
\end{equation}
~\\
but here both $G_i$ and $\Gamma_i$ are fitting constants. We assess the fits using $N=1,2,3$~and~4~exponential functions with the constraint $\sum_{i=1}^N G_i=1$. The values of $G_{1-3}$ and $\Gamma_{1-3}$ for the fit with 3 exponential functions is compared with the values from the lognormal deconvolution with dashed lines in Figs.~\ref{fig:Fitting_noNoise}B-D.

\item The correlation is assumed the form of a stretched exponential:\\
\begin{equation}
c(\tau)=\exp(-2[\Gamma\tau]^{\gamma}), 
\end{equation}
~\\
with $\gamma$ as a stretching coefficient.
\end{enumerate}

The results from the fits are illustrated in Fig.~\ref{fig:Fitting_noNoise} and Table~\ref{tab:Tab1}. Fig.~\ref{fig:Fitting_wNoise} shows the same analysis for the same correlation curve but with added random noise $c_{\epsilon}\in[-0.05,0.05]$ to the original data, \emph{i.e.}~$c(\tau)+c_\epsilon$. The corresponding parameters from the different fitting methods are noted with parenthesis in Table~\ref{tab:Tab1}.

A couple of observations are done with respect to this assessment:\\
\begin{itemize}
\item A simple exponential is not good enough to describe the data.
\item Using 2 exponentials, we already get a very good fit, but it is further improved by using 3 exponentials. There is no improvement of the fit with 4 exponentials.
\item The quality of the fit with 3 exponentials is equally good as the non-regularized CONTIN result ($RP=0$). The results of the three dynamic modes are practically identical for both methods. We could thus qualitatively reproduce the results in the manuscript by just considering a fit with 3 exponentials. However, we find that the CONTIN-approach gives us more flexibility since it does not require a pre-determined number of dynamic modes.
\item The values from the CONTIN fit followed by lognormal deconvolution are not very sensitive to RP. The only issue that is observed is if there is too much smoothing applied (high RP) and the three peaks are thus difficult to distinguish. Here a value of RP$\leq 0.1$ seems to yield results independent of RP.
\item The stretched exponential provides almost as good fit as 3 exponentials and CONTIN, but it is clear that the relaxation rate $\bar{\Gamma}$ does not correspond to the relaxation rates from any of the methods using sums of regular exponentials and is thus also difficult to relate to~\emph{e.g.} an average diffusion coefficient. The main difference with a stretched exponential is that it has an infinite derivative at $\tau=0$ when $\gamma<1$, and thus has a very quick initial decay compared to sums of exponentials. Since there are no indications of such behavior at low $\tau$, we find that the stretched exponential form is not suitable for our study.
\item Neither fitting method shows any significant sensitivity to noisy data, apart from some slight differences in the magnitude of the dynamic modes. In terms of relaxation rates, they provide similar results both with and without noise, which indicates that there are no problems of overfitting the data.
\end{itemize}

\begin{figure}[tbp]
\centering
\includegraphics[width=0.8\textwidth]{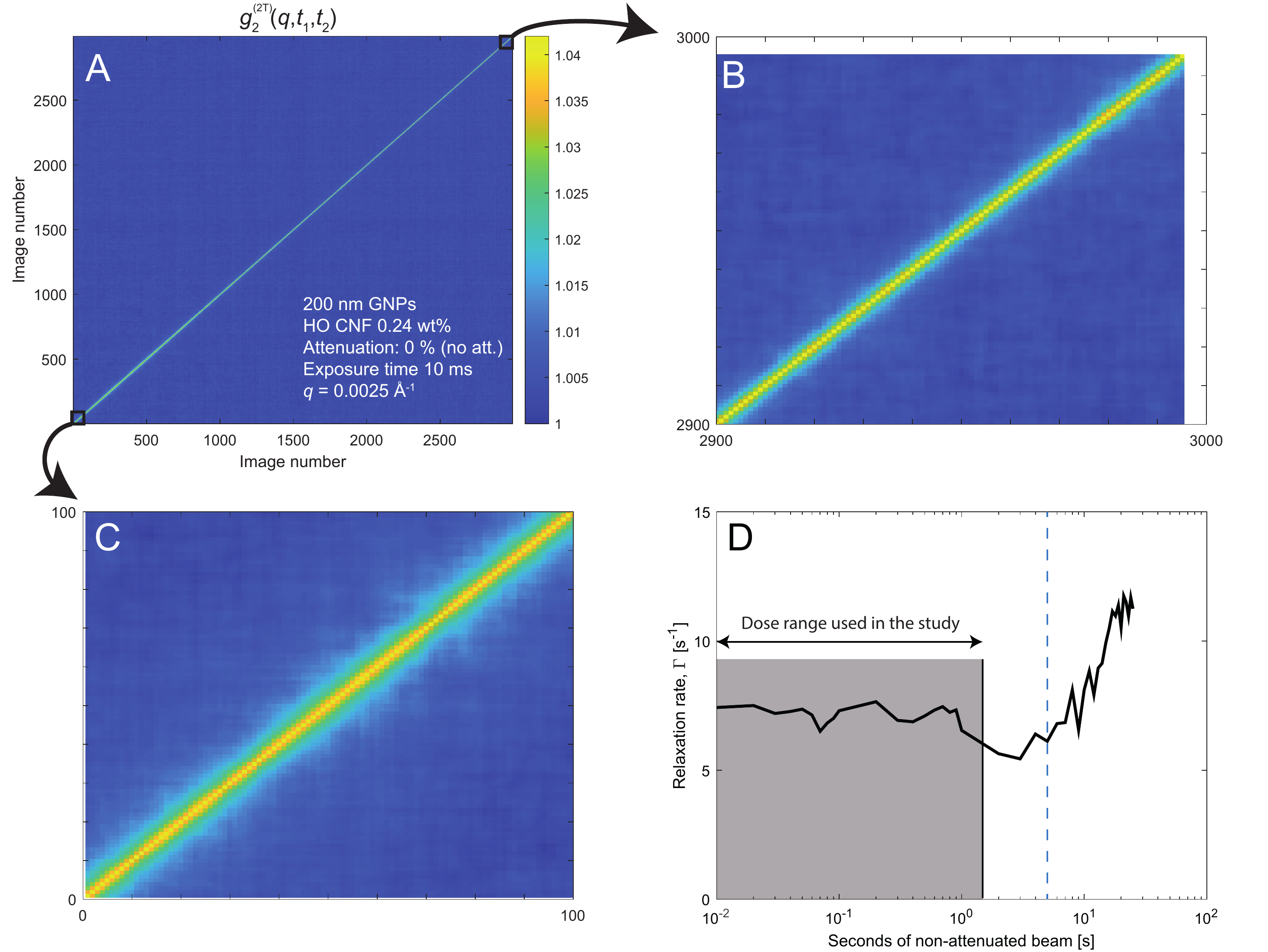}
\caption{Study to assess the influence of the X-ray radiation on the system dynamics in a system of 200~nm GNPs in HO CNF at 0.24~wt\%; (A) the two-time correlation $g_2^{(2T)}(q,t_1,t_2)$ of speckle intensities at $q=0.0025$~\AA$^{-1}$ between images at times $t_1$ and $t_2$ provided as image numbers; (B) zoom-in at the end of the experiment (last 1~s of radiation exposure); (C) zoom-in at the beginning of the experiment (first 1~s of radiation exposure); (D) relaxation rate $\Gamma$ as function of seconds of radiation dose.}
\label{fig:DoseAnalysis}
\end{figure}

\section{Influence of X-ray radiation during experiments}
To study the influence of X-ray radiation on the dynamics, we perform a long-time experiment on the system of 200~nm GNPs in HO CNF at 0.24~wt\%. Here, the experiment is performed with no attenuation to the beam using an exposure time of 10~ms and measuring 3000 subsequent images (30 seconds). To analyze the time-dependent dynamics, we use the two-time correlation function $g_2^{(2T)}(q,t_1,t_2)$ at $q=0.0025$~\AA$^{-1}$, which can provide us directly with the autocorrelations between two detector images at time $t_1$ and $t_2$ (see Fig.~\ref{fig:DoseAnalysis}A-C). From this, we can derive correlation curves at various starting times $t_0$ through $c(\tau)=(g_2(q,t_0,t_0+\tau)-1)/\beta$, which is fitted to a simple exponential function $c(\tau) = \exp(-2\Gamma(t_0)\tau)$. This provides us with a relaxation rate dependent on the experimental time, $\emph{i.e.}$~dependent on the total exposure to the non-attenuated beam, illustrated in Fig.~\ref{fig:DoseAnalysis}D.

The main influence of the radiation is a significant increase in relaxation rates above 5~s of exposure to the non-attenuated beam (above vertical dashed line in Fig.~\ref{fig:DoseAnalysis}D). In the present study, the maximum dose is 1.3~s of exposure to the non-attenuated beam, which seems to have negligible effect on the system dynamics (shaded region) as the relaxation rate is fairly constant.

\begin{figure}[tbp]
\centering
\includegraphics[width=0.99\textwidth]{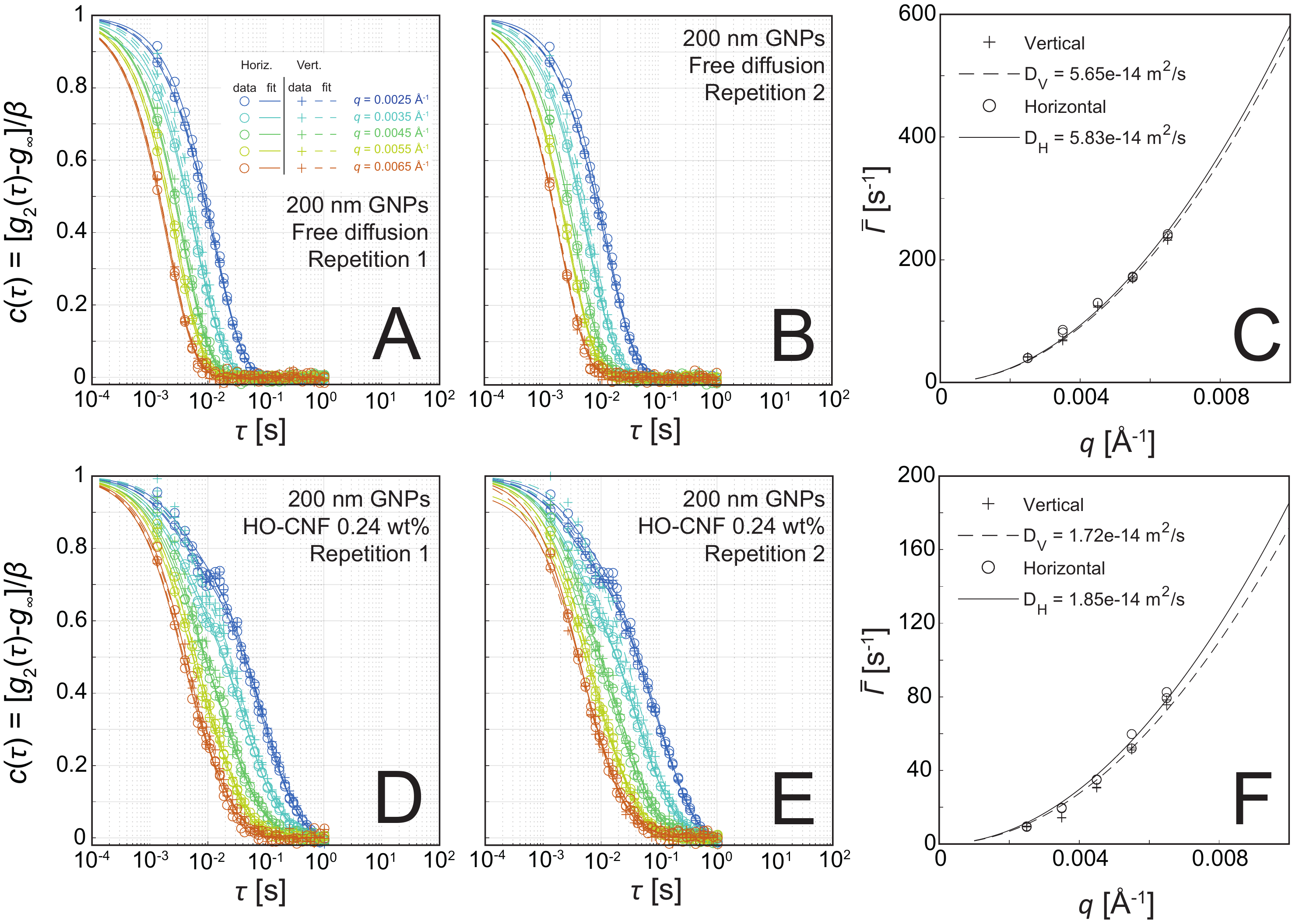}
\caption{Assessment of direction dependent dynamics due to gravity-induced sedimentation; (A) and (B) the autocorrelation curves $c(\tau)$ for 200~nm GNPs freely diffusing in the solvent in two repeated experiments; the circles and crosses show horizontal and vertical data, respectively, with CONTIN fits provided with the solid and dashed curves; (C) the corresponding relaxation rates $\bar{\Gamma}$ in horizontal and vertical directions with corresponding fits of $\bar{\Gamma}=Dq^2$ to obtain diffusion coefficients in each direction; (D)-(F) the same assessment for 200~nm GNPs in 0.24~wt\% HO CNF.}
\label{fig:Sedimentation}
\end{figure}

\section{Influence of gravity during experiments}
To be able to exclude any gravity-induced effects, we divide the detector into four 90$^\circ$ sectors to separate horizontal data (left and right sector on the detector) with vertical data (upper and lower sector on the detector). The square first order auto-correlations $c(\tau)$ for two repeated experiments of freely diffusing 200~nm GNPs are shown in Figs.~\ref{fig:Sedimentation}A-B, where the data is separated into vertical and horizontal directions (crosses and circles in figure). Using CONTIN fits to extract the mean relaxation rate $\bar{\Gamma}$ (solid and dashed curves), we obtain the direction-specific diffusion coefficients $D_V$ in vertical direction and $D_H$ in horizontal directions by assuming $\bar{\Gamma} = Dq^2$ (see Fig~\ref{fig:Sedimentation}C). The same is then also done for two repeated experiments of 200~nm GNPs in 0.24~wt\% HO CNF in Figs.~\ref{fig:Sedimentation}D-F.

From these results, we find that there seems to be no significant difference between the results in horizontal and vertical directions and we can thus rule out any effect of gravity.

\begin{figure}[tbp]
\centering
\includegraphics[width=0.8\textwidth]{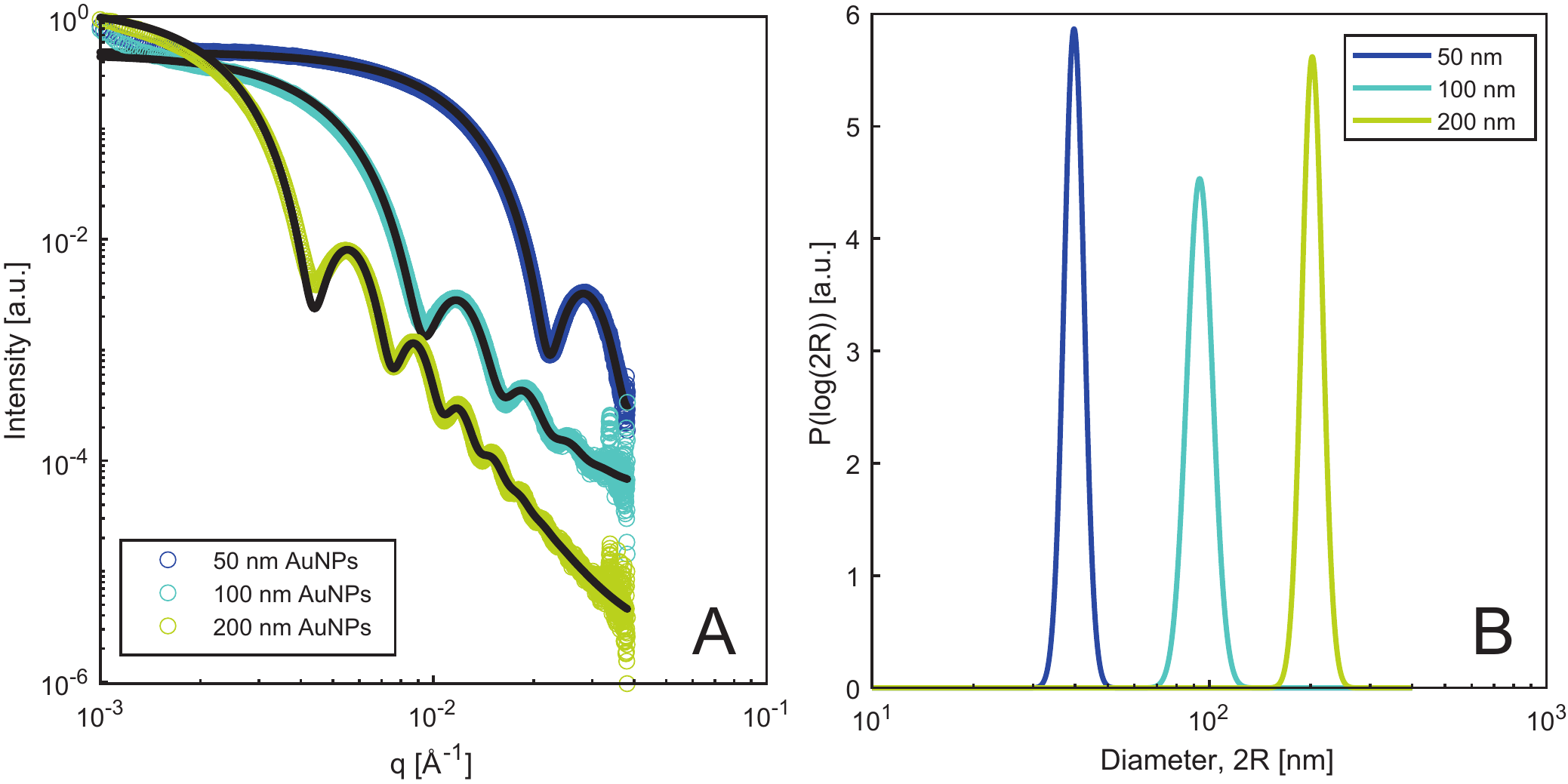}
\caption{Results of SAXS measurements of freely moving GNPs (expected diameters $2R_\text{expected}=50,~100$~and~200~nm) in the solvent; (A) the scattering intensity $I$ as function of wave vector $q$ (symbols) is fitted (black curves) assuming spherical particles with diameter $2R$ with polydispersity described by a log-normal distribution; (B) the resulting size distribution from the SAXS fitting.}
\label{fig:SAXSGNP}
\end{figure}

\begin{figure}[tbp]
\centering
\includegraphics[width=0.99\textwidth]{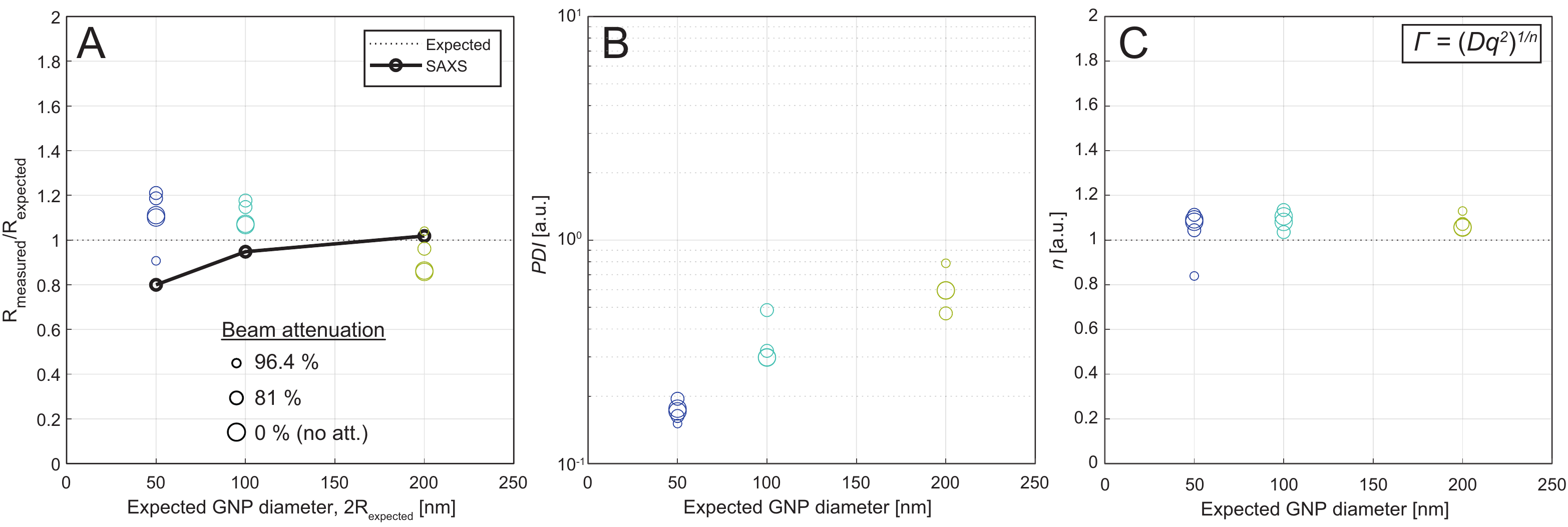}
\caption{Results of XPCS measurements of freely moving GNPs (expected diameters $2R_\text{expected}=50,~100$~and~200~nm) in the solvent; (A) the geometric mean radius from SAXS compared to the hydrodynamic radius from XPCS; (B) polydispersity index $PDI$; (C) dynamic scaling exponent (assuming $\Gamma=(Dq^2)^{1/n}$).}
\label{fig:XPCSGNP}
\end{figure}

\begin{table}[tbp]
\begin{tabular}{l|l|l|l|}
         	      & 50 nm & 100 nm & 200 nm \\
         	      \hline
         	      \hline
2$R_{SAXS}$ [nm] &      39.9     &     94.8        &   203.6   \\
\hline
$PD_{SAXS}$ [a.u.] &      0.068     &  0.088    &    0.071    \\
\hline
\hline
$D_0$ [$10^{-14}$ m$^2$/s]   &  17.5  	&    9.04       &       5.65            \\
\hline
2$R_{XPCS}$ [nm] &    57.5   &  111   &    178    \\
\hline
  $PDI_{0}$ [a.u.] &   0.176   &  0.351      &  0.554     \\
\hline   
\end{tabular}
\caption{Table of parameters from SAXS and XPCS measurements of freely moving GNPs in the solvent.}
\label{tab:TabGNP}
\end{table}

\section{Sizes and dynamics of GNPs in pure solvent}
The static and dynamic properties of the GNPs were studied in pure dispersions of propylene glycol (PG) using both the static (SAXS) and dynamic (XPCS) data from the experiments. In Fig.~\ref{fig:SAXSGNP}, we find the SAXS curves of the three GNP sizes with circle symbols. The data is fitted in SasView using a sphere-model with lognormal polydispersity~\cite{SasView}. The resulting distributions are illustrated in Fig.~\ref{fig:SAXSGNP}B with the values of mean radius $R_{SAXS}$ and polydispersity $PD_{SAXS}$ given in Table~\ref{tab:TabGNP}. Here we can see that the geometric size of the smaller particles is smaller than the expected diameter of 50~nm. 

The dynamics of the freely moving GNPs in the solvent was measured with XPCS and the Brownian diffusion coefficient $D_B$ was extracted according to the procedure in the manuscript. Through the Stokes-Einstein relationship $D_B= k_B T/(6\pi\eta R_{H})$ (temperature $T=298$~K and viscosity of PG determined by Sagdeev~\emph{et al.}\cite{sagdeev2017density}), the diffusion coefficient is used to determine the hydrodynamic radius $R_{H}$ for different attenuation levels as illustrated in Fig.~\ref{fig:XPCSGNP}A. 

Firstly, it should be noted that the comparison with the expected value (assumed at $T=298$~K) is highly dependent on an accurate measurement of the effective temperature in the beam, which was not possible to do in the experiment. Just a 1-2~$^\circ$C error in the effective temperature can cause a 10~\% difference between hydrodynamic radii due to the temperature dependence of the solvent viscosity. This is the likely cause for the apparent decrease of hydrodynamic radii with a stronger dose as it is likely that a 5~times stronger beam could result in a slight local heating of the irradiated sample. It should also be added that a 5~times smaller dose adds more noise to the data and larger error to the extracted diffusion coefficient. The error is also larger for the smaller particles, owing to their faster dynamics and therefore less usable autocorrelation data at higher $q$.

Despite these uncertainties, there are some interesting trends when comparing the geometric and hydrodynamic radii. The geometric radius is lower than the expected value for smaller GNPs but close to the expected size for the larger ones. On the other hand the hydrodynamic radius has a decreasing trend and a larger difference to the geometric radius for smaller GNPs. This is likely due to the polymer coating of approximately 1-5~nm covering the particle for steric stabilization, causing the particle to move slower than expected from its geometric size. This is naturally more prominent for the smallest particles as the relative thickness of the polymer layer is smaller for the larger GNPs. 

The polydispersity index ($PDI$) describing the non-uniformity of dynamics in the systems is illustrated in Fig.~\ref{fig:XPCSGNP}B and determined with expressions in the main manuscript. We find slightly higher $PDI$ for the larger GNPs, but they still show fairly high uniformity of dynamics with values of $PDI<1$. Note that the values of $PDI$ cannot be directly compared with the $PD_{SAXS}$ as they have different definitions. The dynamic scaling exponent assuming $\Gamma=(Dq^2)^{1/n}$ is illustrated in Fig.~\ref{fig:XPCSGNP}C, where values are close to (albeit a bit larger than)~1 and thus little anomalous diffusion in the systems, \emph{i.e.} the GNPs have a mean square displacement that is linearly proportional to time.

The mean of the data at 0\% and 81\% attenuation levels is used as the reference values $D_0$ (related to mean hydrodynamic radius $R_{XPCS}$) and $PDI_0$ in the main manuscript and summarized in Table.~\ref{tab:TabGNP}.

\begin{figure}[p]
\centering
\includegraphics[width=0.55\textwidth]{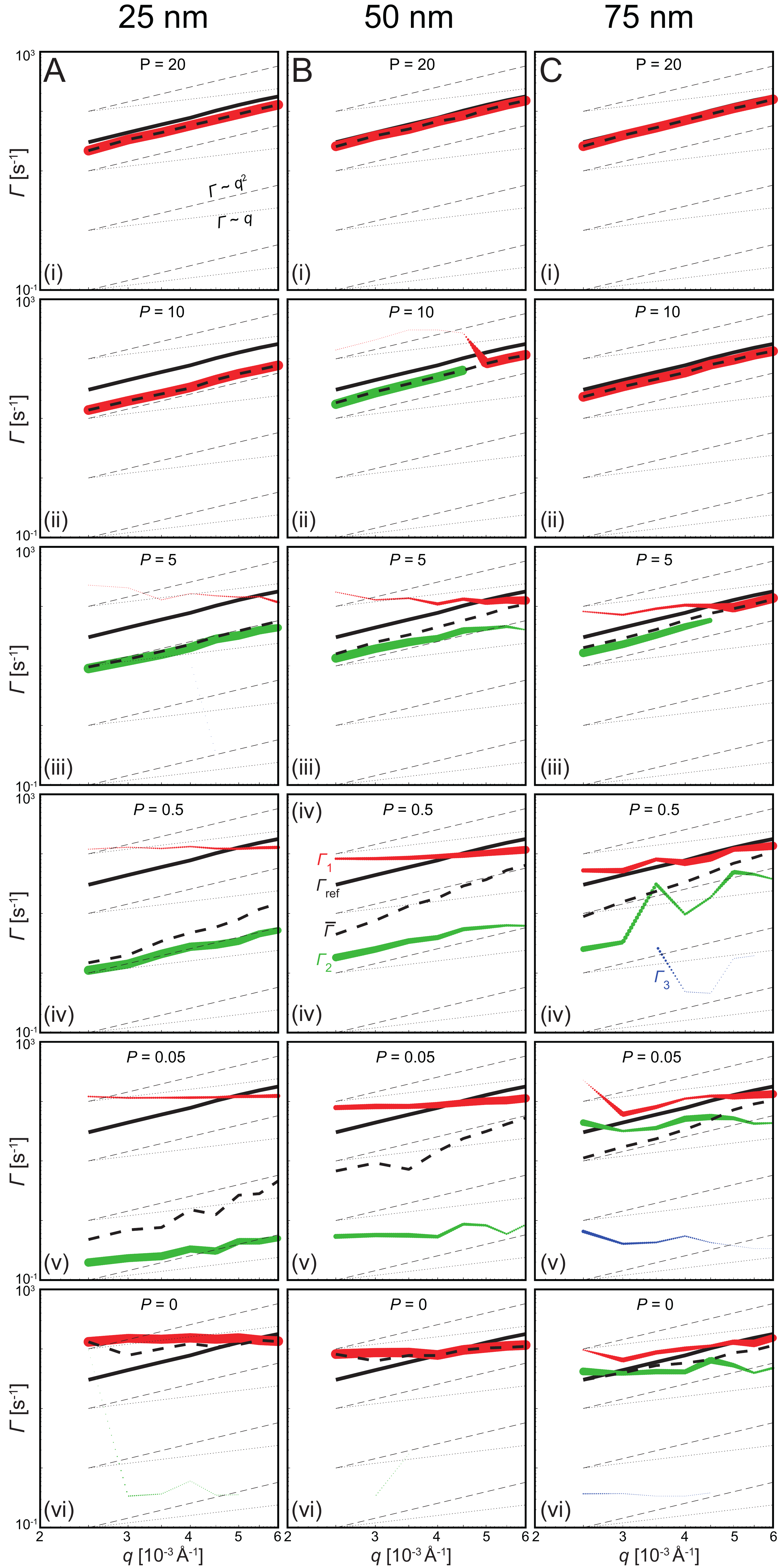}
\caption{Results from the numerical simulations of confined Brownian motion of 200~nm spherical nanoparticles; ; the evolution of the relaxation rates $\Gamma_1$, $\Gamma_2$ and $\Gamma_3$ as function of $q$ are drawn with the line width scaled with the magnitude of the modes and linear interpolation between the $q$-values; the thin dashed and dotted lines indicate expected scaling behavior for Brownian ($\Gamma\varpropto q^2$) and ballistic ($\Gamma\varpropto q$) dynamics, respectively; the thick black solid and dashed lines show the free diffusion of GNPs ($\Gamma_\text{ref}$) and mean relaxation rate ($\bar{\Gamma}$), respectively;(A)-(C) shows the results using cell sizes of 25, 50 and 75~nm, respectively; panels (i)-(vi) show the results for various values of permeability $P$.}
\label{fig:Simulations}
\end{figure}

\section{Results from the digital twin using various cell sizes}
The same procedure to study evolution of dynamic modes in the simulated system, as described in the main manuscript, was performed for two other cell sizes $25$ and $75$~nm (the results for $50$~nm was shown in the main manuscript). The results are shown in Fig.~\ref{fig:Simulations}, illustrating different transition scenarios for the extracted modes. This indicates that the trends can be matched to the experimental trends to find an average cell size, corresponding to the undisturbed ''wiggle-room'' for the GNPs in the CNF network.

\begin{figure}[tbp]
\centering
\includegraphics[width=0.70\textwidth]{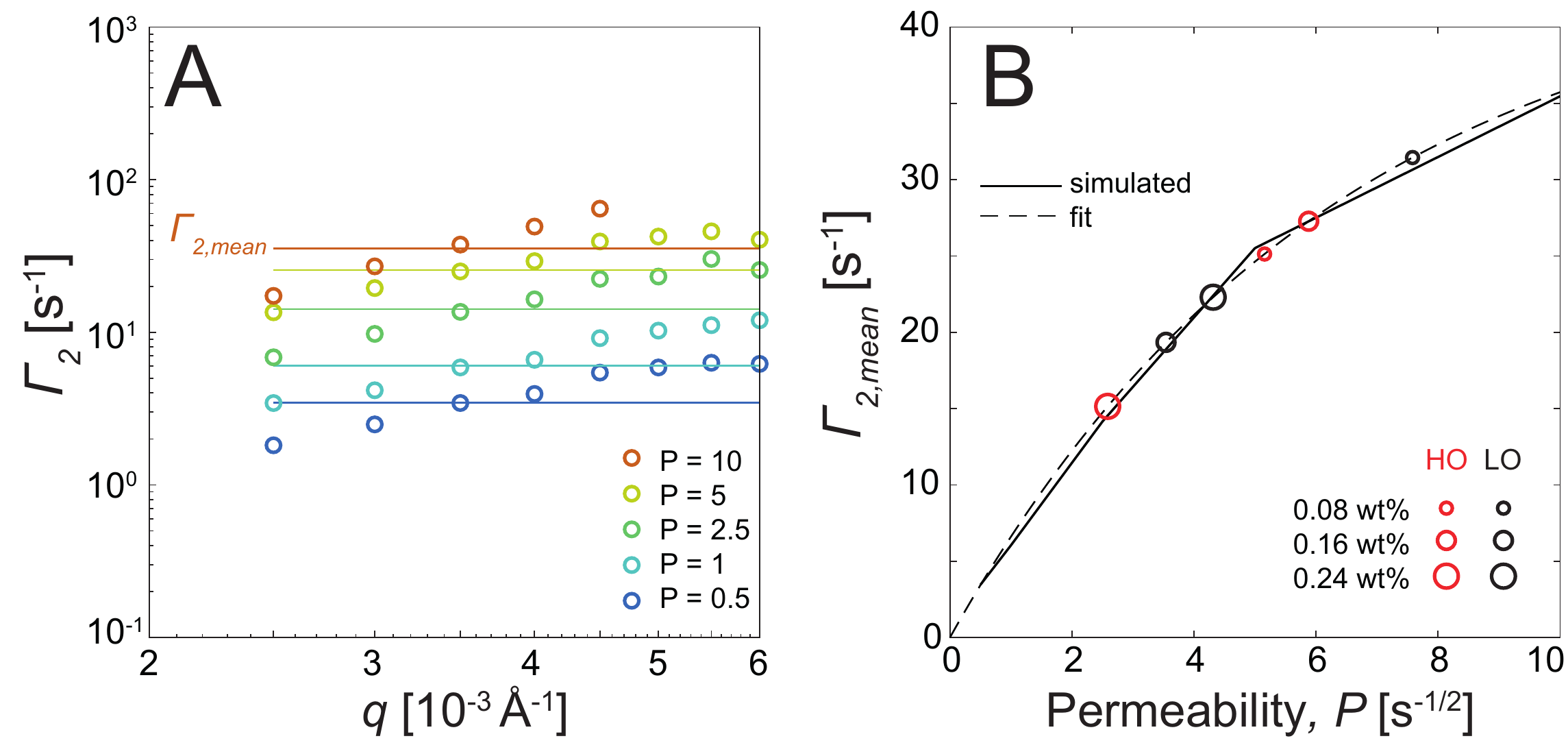}
\caption{Quantification of permeability $P$ in the CNF networks for 200~nm GNPs using the digital twin to compare relaxation rates $\Gamma_{2,\text{mean}}$; (A) Relaxation rate of second dynamic mode $\Gamma_2$ as function of $q$ at various permeabilities from the digital twin; (B) the weighted mean $\Gamma_{2,\text{mean}}$ (illustrated with solid lines in A) as function of permeability $P$ is fitted and used to obtain values of $P$ for the experimental systems.}
\label{fig:DigTwinQuant}
\end{figure}

\section{Quantification of permeability using the digital twin}
As observed from the results in the digital twin in Fig.~\ref{fig:Simulations}B, the dynamic modes in (ii)-(iv) corresponding to permeabilities between $P=0.5$ and 10 for 50~nm cell size match the experimental trends qualitatively (Fig.~4 in main manuscript). The characteristic $P$-dependent feature is the decrease of the mean relaxation rate of the second dynamic mode $\Gamma_2(q)$ with decreasing $P$. Using the mean relaxation rate $\Gamma_{2,\text{mean}}$ (mean taken over $q$ weighted with magnitude $G_2(q)$) from the digital twin at various $P$, illustrated in Fig.~\ref{fig:DigTwinQuant}A, we fit an exponential function $\Gamma_{2,\text{mean}}(P) = C_1[\exp(-C_2 P) -1]$ illustrated with dashed line in Fig.~\ref{fig:DigTwinQuant}B. By extracting the same quantity $\Gamma_{2,\text{mean}}$ from the experimental systems we thus can use this function to find the permeability $P$ of the CNF networks (values in Fig.~6 in main manuscript).

\clearpage
\bibliography{SI_XPCS-paper}